\documentclass[%
 reprint,
nofootinbib,
 amsmath,amssymb,
 aps,
]{revtex4-2}

\usepackage{graphicx}
\usepackage{dcolumn}
\usepackage{bm}
\usepackage{hyperref}

\usepackage[plainruled]{algorithm2e}
\usepackage{xcolor}
\usepackage{braket}
\usepackage{amsthm}
\usepackage[caption=false]{subfig}
\usepackage{calc}

\definecolor{lightblue}{RGB}{73,151,208}
\definecolor{crimson}{RGB}{140,41,53}

\newcommand*{\mask}[2]{%
	\mathord{\makebox[\widthof{\(#1\)}]{\(#2\)}}%
}

\hypersetup{
    colorlinks,
    linkcolor={crimson},
    citecolor={lightblue},
    urlcolor={lightblue}
}

\makeatletter
	\def\@algocf@capt@plainruled{above}
	\renewcommand{\algocf@caption@plainruled}{%
	\vskip\AlCapSkip%
	\box\algocf@capbox%
	\vskip 5\algoheightrule}%
\makeatother

	\makeatletter
	\def\env@cases{%
	  \let\@ifnextchar\new@ifnextchar
	  \left\lbrace
	  \def\arraystretch{1.2}%
	  \array{l@{\quad}l@{}}%
	}
	\makeatother

\renewcommand{\arraystretch}{1.5}

\newtheorem{definition}{Definition}[section]
\newtheorem{lemma}{Lemma}[section]
\newtheorem{theorem}{Theorem}[section]
\newtheorem{problem}{Problem}[section]
\newtheorem{corollary}{Corollary}[theorem]
\newtheorem{remark}{Remark}[section]

\begin{document}

\preprint{APS/123-QED}

\title{Quantum Hypothesis Testing with Group Structure}


\author{Zane M. Rossi}%
\email{zmr@mit.edu}
\author{Isaac L. Chuang}%
\affiliation{%
    Department of Physics, Center for Ultracold Atoms, and Research Laboratory of Electronics\\Massachusetts Institute of Technology (MIT), Cambridge, Massachusetts 02139, USA
}%

\date{\today}

\begin{abstract}
    The problem of discriminating between many quantum channels with certainty is analyzed under the assumption of prior knowledge of algebraic relations among possible channels.
	It is shown, by explicit construction of a novel family of quantum algorithms, that when the set of possible channels faithfully represents a finite subgroup of SU(2) (e.g., $C_n, D_{2n}, A_4, S_4, A_5$) the recently-developed techniques of quantum signal processing can be modified to constitute subroutines for quantum hypothesis testing.
	These algorithms, for group quantum hypothesis testing (G-QHT), intuitively encode discrete properties of the channel set in SU(2) and improve query complexity at least quadratically in $n$, the size of the channel set and group, compared to na\"ive repetition of binary hypothesis testing.
	Intriguingly, performance is completely defined by explicit group homomorphisms; these in turn inform simple constraints on polynomials embedded in unitary matrices. These constructions demonstrate a flexible technique for mapping questions in quantum inference to the well-understood subfields of functional approximation and discrete algebra.
	Extensions to larger groups and noisy settings are discussed, as well as paths by which improved protocols for quantum hypothesis testing against structured channel sets have application in the transmission of reference frames, proofs of security in quantum cryptography, and algorithms for property testing.
\end{abstract}


\maketitle


\section{Introduction} \label{section:introduction}

		Hypothesis testing is a fundamental statistical method with wide application in classical and quantum contexts. Seminal work \cite{helstrom} has led to a deep information-theoretic understanding of binary hypothesis testing for \emph{quantum states}, but only quite recently have analogous lower bounds been proven for error in discrimination among \emph{quantum channels} \cite{pirandola}. This forty-year gap between mature theories for quantum hypothesis testing (QHT), realized as quantum state and channel discrimination respectively, follows from the far richer structure of the latter problem. I.e., general quantum channel discrimination protocols may be adaptive, entanglement-assisted, and use auxiliary qubits; moreover, the concomitant optimizations over (possibly adaptive) preparations and measurements are computationally expensive.

		It is known that sharpening the problem of quantum channel discrimination to narrower settings can drastically alter algorithmic efficiency, the requirement of entanglement, the requirement of auxiliary qubits, and the ease of both theoretical and computational analysis \cite{acin, duan, duan_feng_ying}. This work considers one such narrower statement of QHT for discriminating quantum channels.
    
	\subsection{Problem statement} \label{subsection:problem_statement}

		We state our problem as a game. Consider a party with access to a small (single-qubit) quantum computer; she is able to apply unitary operations of her choice to this qubit, measure this qubit in chosen bases, and store the resulting classical data for as long as she likes, possibly using this information to instruct future actions. She is furthermore permitted query access to an oracle whose result is the application of a single-qubit unitary quantum channel $\mathcal{E}_i$. This channel is from a publicly known set $S$ (hereafter the \emph{query set}) of $n$ distinct unitary channels. Queries consistently apply $\mathcal{E}_i$, and $i$ is unknown.

		\begin{problem} \label{s_qht_problem}
		    An \emph{S-QHT Problem} is any instance wherein a party given access to $\mathcal{E}_i$ for unknown $i \in [n]$ is tasked with the following: in as few queries as possible determine, with certainty, the hidden index $i$. The \emph{minimal expected query complexity} the party is able to achieve is denoted $q_s$ and is taken over an assumption of equal priors on $\{\mathcal{E}_\ell\}_{\ell \in [n]} = S$, a set of distinct single-qubit unitary quantum channels.
		\end{problem}
		
		The prefix \emph{S} in Problem \ref{s_qht_problem} denotes QHT with respect to a \emph{set} of quantum channels. This work examines only specific subsets of S-QHT games. Moreover, this work considers a specific resource model, described informally at the beginning of this section and depicted in Figure \ref{figure:serial_adaptive}.

		As described in Subsection \ref{subsection:prior_work}, na\"ive upper and lower bounds on $q_s$, even for general $S$, can be computed without difficulty, although the gap between these bounds is in general large, i.e., exponential in the instance size $\lvert S \rvert$ \cite{duan_feng_ying}. A primary interest is thus to derive a set of properties on the set $S$ for which a lower bound for $q_s$ \emph{dependent on the structure of $S$} can be both (1) proven and (2) asymptotically achieved by a quantum algorithm \emph{exploiting the structure} of $S$ to generate a strategy for playing an instance of S-QHT (Problem \ref{s_qht_problem}).

		This work provides one such sufficient condition on $S$. These constraints not only enable proof of query complexity lower bounds and constructions of algorithms achieving these bounds, but permit the cross-application of diverse methods in abstract algebra and functional approximation theory to quantum information and inference. This work considers the specific constraint that $S$ additionally faithfully represents a finite subgroup $G < \text{SU(2)}$ (i.e., it is a representation of a finite subgroup of the group of single-qubit unitary transformations).

			\begin{definition} \label{def:faithful_rep}
				A channel set $S$ is said to \emph{faithfully represent} a finite group $G$ if the elements of $S$ have the form $\{U_g\}_{g\in G}$ such that, respecting some natural product operation for elements in $S$, $U_g U_h = U_{gh}$ for $g, h \in G$, and moreover that the group homomorphism $g \mapsto U_g$ is injective, ensuring $\lvert S \rvert = \lvert G \rvert$.
			\end{definition}

		A variant of S-QHT incorporating the condition discussed above is denoted by G-QHT (Problem \ref{g_qht_problem}). While this work considers groups $G < \text{SU(2)}$, this game naturally extends to finite representations embedded in any Hilbert space.

			\begin{problem} \label{g_qht_problem}
				An instance of Problem \ref{s_qht_problem} with the additional constraint that $S$ faithfully represents a finite group $G$ is an instance of a \emph{G-QHT problem} or \emph{G-QHT game}.
			\end{problem}

		Before discussing this new game further, it is worthwhile to understand previous results in unitary quantum channel discrimination, to which these games have non-trivial relation. These results support why one should expect that the family of sets $S$ which obeys the properties of Lemma \ref{def:faithful_rep} is rich enough to furnish non-trivial instances of QHT, and why even in a limited resource model algorithms to solve G-QHT efficiently exist.
    
	\subsection{Prior work} \label{subsection:prior_work}

		The problem of \emph{binary} quantum channel discrimination is well-understood under the assumption that the set of possible channels, i.e., the \emph{query set}, denoted $S$, comprises only unitary channels. Foundational work by Ac\'in \cite{acin} asserts that there is always some finite upper bound\footnote{This furnishes a loose upper bound for multiple unitary channel discrimination as well; one performs perfect discrimination on pairs of elements in $S$, eliminating channels one by one; this is the \emph{standard reduction} to binary QHT.} on $q_s$ for achieving \emph{perfect discrimination} for any finite $S$ with distinct, known, unitary elements. Moreover it is known that \emph{in the binary case}, under the assumption that the discriminating party may apply unitary operations of their choice, neither entanglement nor auxiliary systems nor adaptive protocols are required to achieve optimal query complexity \cite{duan, duan_feng_ying}.

		For \emph{binary discrimination} among pairs of \emph{general quantum channels}, necessary and sufficient conditions are known for the achievability of perfect quantum channel discrimination in terms of the channel's respective Choi matrices \cite{pirandola}. Moreover, various \emph{general lower bounds} are known for the symmetric error of discrimination (given a fixed number of channel uses) for binary and multiple quantum channel discrimination, as well as some conditions on the set $S$, e.g., teleportation-covariance (telecovariance) and geometric uniform symmetry (GUS) under which these bounds can be improved upon and, in the former, more restrictive setting of telecovariance, asymptotically achieved \cite{pirandola, zhuang_pirandola}. Such simplifying conditions have also been studied in the multiple unitary channel case for group covarianct query sets for \emph{non-adaptive} quantum strategies \cite{hashimoto_10}.

		While it is known that entanglement (and in fact any resource in a convex resource theory like quantum mechanics \cite{takagi, takagi_general}) can be useful in quantum hypothesis testing among non-unitary channels, the performance of entanglement-free or low-entanglement strategies for multiple quantum channel discrimination remains largely unstudied, even in its simplest, unitary form. Namely, while intriguing examples for methods of discrimination among large sets of unitary operators where the use of entanglement improves query complexity have been given \cite{duan_feng_ying}, the necessity of entanglement is not known. Moreover, the power afforded to quantum hypothesis testing strategies for quantum channels using entanglement \emph{and which are also adaptive} has been shown to be non-trivial in the case of non-unitary channels, where even adaptiveness alone may assist algorithmic performance \cite{harrow-10, sacchi_05}.

		Many of the techniques referenced above are agnostic to the structure of $S$; however, the notion that the structure of the query set should inform the structure of optimal procedures to differentiate members of $S$ is an old and clever idea, and indeed can provide optimal hypothesis testing protocols for query sets comprising \emph{quantum states} which are group covariant \cite{davies}. It is as a generalization of this setting to quantum channels that Problem \ref{g_qht_problem} (G-QHT) finds its form. Moreover, the study of discrete and especially non-abelian algebraic objects in the context of quantum information is not new, and underlies many open problems, e.g., the dihedral hidden subgroup problem \cite{kuperberg} and its reductions to various lattice problems \cite{regev}, as well as the symmetric hidden subgroup problem and its reductions to graph isomorphism \cite{childs_10}.

		Multiple hypothesis testing for quantum channels is not merely of independent quantum-information-theoretic interest either, but has found use in designing protocols for the optimal transmission of reference frames \cite{chiribella_05} (i.e., when the query set is a compact group and the aim is estimation of a fixed unitary transformation). Discretized versions of this problem also naturally connect to the study of group frames and SIC-POVMs \cite{group_frames, frame_introduction}, e.g., as discussed in Lemma \ref{lemma:discrete_reference_frame}.
		
		While left as an open extension to this work, application of methods for quantum hypothesis testing against quantum channels where the $n$-th channel application depends non-trivially on the previous $n-1$ applications, i.e., \emph{memory channels} \cite{memory_channels_08} also have application to proofs of the general impossibility of quantum bit-commitment \cite{bit_commitment_07}, and are of interest in quantum cryptography.

		In what follows we more concretely define our algorithmic resource model, provide an example of why it might be expected that the question of achievability within the exponential gap between the na\"ive upper and lower bounds on query complexity for multiple quantum hypothesis testing is richly structured, and finally give an outline for the methods of proof employed in analyzing this structure.
    
	\subsection{Our approach}

\begin{figure*}[htbp!]
				\begin{center}
				    \includegraphics[width=0.75\textwidth]{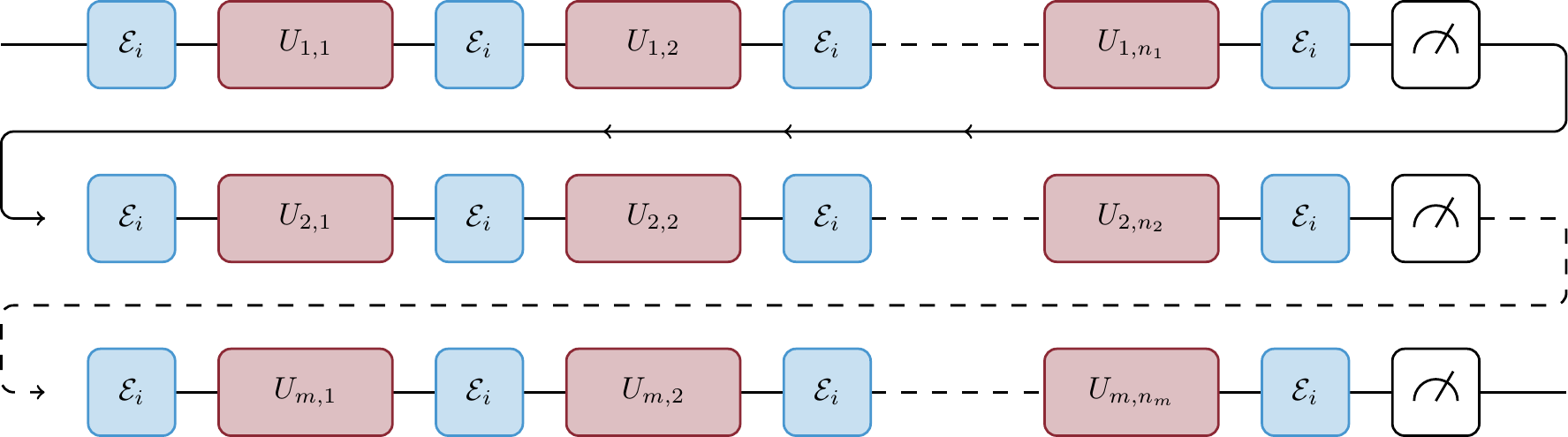}
				\end{center}
	  			\caption{A general circuit to perform QHT in the \emph{serial adaptive query model}. The unitary operators $U_{i, j}$ for $i \in [n_j], j \in [m]$ may depend on previous single-qubit projective measurements $\Lambda_k$ for $k < j$, for $j \in [m]$, communicated by stored classical bit strings of reasonable finite length (represented by arrows). Each row in the figure is a quantum circuit applied to a qubit prepared from classical information depending only on the results of previous measurements. The serial nature of the discrimination protocol to determine the unknown channel is evident; when the protocol terminates a known classical function is computed on the set of measurement results (here, a bit-string of length $m$), equivalently $\Lambda_k$ for $k \in [m]$, to infer the hidden channel. Other models one can consider are discussed in Figure \ref{figure:circuit_geometry}.}
	  			\label{figure:serial_adaptive}
			\end{figure*}

        The statement of G-QHT (Problem \ref{g_qht_problem}) together with the serial adaptive query model depicted in Figure \ref{figure:serial_adaptive} raises the question of whether this model is (1) interesting, (2) non-trivial, and (3) tractable to analyze; this section addresses these questions.

		The player challenged in G-QHT to determine the hidden index $i$ of the queried channel $\mathcal{E}_i$ is afforded precious few quantum resources. Stating it another way, the player is forced to devise quantum strategies in the \emph{serial adaptive query model}. In this model, pictured in Figure \ref{figure:serial_adaptive}, the player may only intersperse their oracle queries with measurements and unitary operations depending on previous measurements. Serially, the querent learns progressively more about the hidden index $i$, adaptively modifying her approach. Under the assumption of a small quantum computer and a reasonable classical one, this is the most general approach she may take, assuming all measurements are projective and she wishes to determine $i$ with certainty. Furthermore, in this model, query complexity is a reasonable metric by which to judge algorithmic performance.

\begin{figure*}[htbp!]
				\begin{center}
					\includegraphics[width=0.75\textwidth]{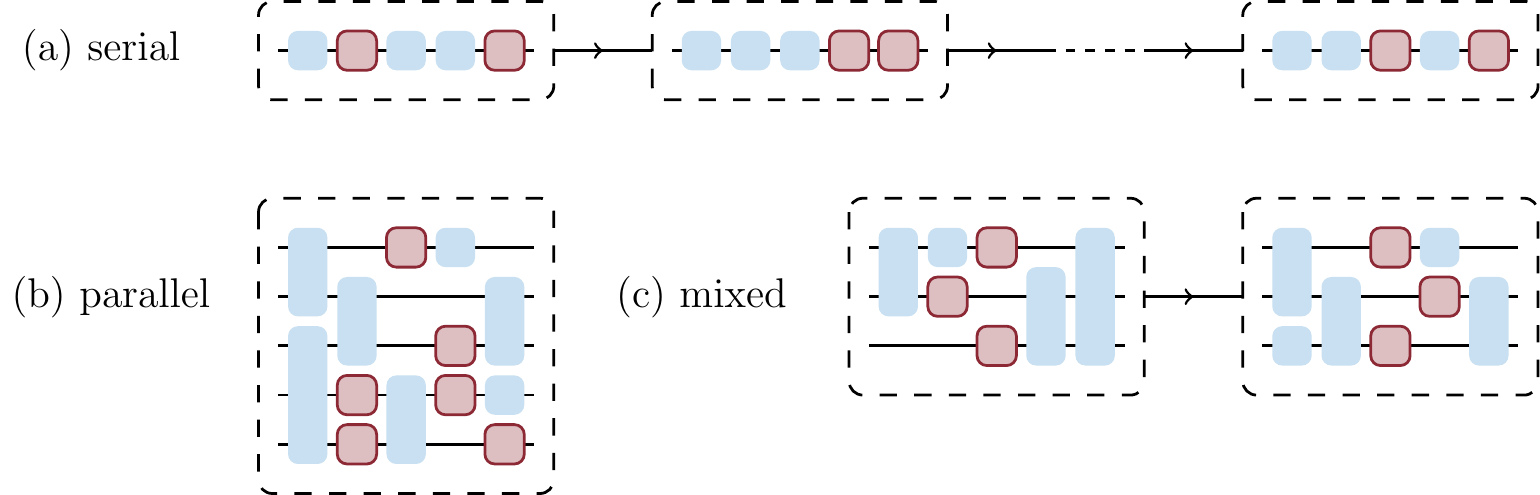}
				\end{center}
	  			\caption{Simplified illustrations of different models for quantum circuits performing QHT. Depicted are (a) serial adaptive, (b) parallel, and (c) mixed strategies. Given query access to a finite number of applications of the unknown quantum channel (red, outlined), in the figure 6 applications, the querent is conferred the ability to perform unitary operations (blue, non-outlined) of her choice. Blue operations shown are arbitrarily structured and for demonstrative purposes only. For serial adaptive strategies, (a), dashed boxes indicate regions between which only classical information is transmitted (i.e., measurement results). In (b) and (c) entanglement, auxiliary qubits, and collective measurements can, in general, improve the performance of QHT algorithms. Preparations and measurements are not explicitly shown.}
	  			\label{figure:circuit_geometry}
			\end{figure*}

		In addition to the serial adaptive query model, we can quickly chart algorithmic schemes for instances of G-QHT where the querent is afforded a larger quantum computer. In this case, the possibility for multiple-qubit\footnote{One could of course also imagine access to qudits, or indeed stranger Hilbert spaces.} unitaries and collective measurements gives rise to a variety of series, parallel, and mixed strategies, which may be adaptive or non-adaptive. The relative discriminating power of these models for specific instances of QHT and specific query sets is not wholly understood. An informal depiction of some of these models is give in Figure \ref{figure:circuit_geometry}.

		As the querent in the course of playing the G-QHT game is allowed to store reasonable amounts of classical information, all that is asked of a successful quantum algorithm for G-QHT in the serial adaptive query model is that it is able to decide the hidden index $i$ according to some efficiently computable function on any of its \emph{probable} binary qubit measurement outputs. This statement is made concrete in Definition \ref{def:serial_adaptive}.

    		\begin{definition} \label{def:serial_adaptive}
    			A quantum algorithm in the serial adaptive query model is said to decide on a query set $S$ of distinct unitary quantum channels of size $n$ in $q_s$ queries if there exists, for all $i \in [n]$ a computable deterministic function $f: \{0, 1\}^{m} \rightarrow [n]$ that returns the hidden index $i$ with certainty, on all probable (i.e., non-zero probability outcomes of) $m$ projective single-qubit measurements $\{\Lambda_{\ell}\}_{\ell \in [m]}$ resulting from the action of $\mathcal{E}_i$ in a serial adaptive protocol defined by the quantum algorithm that uses $q_s$ oracle queries. This definition can be suitably modified replacing $S$ with $G$, a faithful representation of the group $G$ in a specified Hilbert space.
    		\end{definition}

		While we will soon be interested in the efficiency of a single-qubit serial adaptive query model algorithm in deciding a set $S$ which faithfully represents a finite subgroup $G < \text{SU(2)}$, and indeed whether, for these special sets, query-complexity-optimal, entanglement-free, serial adaptive protocols similar to those constructed in \cite{duan} are possible to construct, it is worthwhile to look at a simple, concrete instance of our game, and the function $f$ it induces according to Definition \ref{def:serial_adaptive}.

		We introduce a minimal instance of G-QHT which, in addition to demonstrating why the na\"ive upper bounds on query complexity discussed in Subsection \ref{subsection:prior_work} are in general not tight, also captures some of the intuitive motivations for the major results of this work for more complicated query sets. The following example has the added benefit of (1) requiring no explicit mention of QSP (Section \ref{section:qsp_overview}) in its construction and proof of optimality, and (2) providing some intuition for why QSP is natural to call on to solve the shortcomings that emerge in applying the strategy of Lemma \ref{lemma:simple_cyclic_group} to more general query sets.

			\begin{lemma} \label{lemma:simple_cyclic_group}
				For natural numbers $n$ there exists a quantum algorithm in the serial adaptive query model that perfectly decides any channel set $S$ that faithfully represents a cyclic subgroup $C_{2^n} < SU(2)$, and which requires $2^n - 1$ oracle queries.

				\begin{proof}
					For $C_{2^n}$, group elements are identifiable with binary strings of length $n$ of which there are $2^n$, namely labeling according to the angle of rotation in the Bloch sphere in units of $2^{1 - n}\pi$ such that the queried channel rotates about a known fixed axis by this angle. Concretely, up to overall unitary transformation the query set is
						\begin{equation}
							S = \{R_{x}(m\cdot\pi/2^{n-1})\}, \; m \in [2^n].
						\end{equation}
					Any decision protocol using one qubit for readout can provide at most one bit of information as to the $n$-bit label for the queried group element.\footnote{Note that these don't need to bits in the label of the queried channel, but rather some set of bits which, at the conclusion of the algorithm, can be taken by the function $f$ to the hidden index $i$ deterministically.} We read from least (LSB) to most (MSB) significant bit by the following algorithm:
						\begin{enumerate}
							\item Prepare $\ket{0}$. Query the channel $2^{n - 1}$ times and measure in the standard basis, reading the LSB.

							\item Dependent on the measurement in the previous step the possible query set $S^\prime$ has description
								\begin{align*}
									\{R_{x}(m\cdot\pi/2^{n-2} + \pi/2^{n-1})\} 
									&
									\;\;\text{ if measured $\ket{1}$}\\
									\{R_{x}(m\cdot\pi/2^{n-2})\} 
									& 
									\;\;\text{ if measured $\ket{0}$},
								\end{align*}
							for $m \in [2^{n - 1}]$. The latter is a representation of the cyclic group of order $2^{n - 1}$. The former, if each query is preceded by a unitary $U = R_{x}(-\pi/2^n)$, is also a representation of this cyclic group.
							
							Set $U = R_{x}(-b\cdot\pi/2^{n-1})$, where $\ket{b}$ was measured in the previous step.

							\item Apply $U$ before each of $2^{n-2}$ channel applications to bit-shift the label of the queried group element. Repeat algorithm for a cyclic group of size $2^{n-1}$.
						\end{enumerate}
					For the cyclic group of order 2, consisting of the identity channel and a $\pi$-rotation, the decision protocol is obvious. By recursion, the total decision protocol has query complexity $ 2^{n-1} + 2^{n-2} + \cdots  + 1 = 2^{n} - 1.$ Optimality follows from the optimality of phase estimation.
				\end{proof}
			\end{lemma}

		The methods used in the proof of Lemma \ref{lemma:simple_cyclic_group} illustrate an important concept: if the query set $S$ is highly structured, binary measurement results can effectively correspond to halving the size of the remaining search space (or equivalently excluding, with one measurement, half of the possible channels). Here, compared to the upper bound given by the standard reduction to binary QHT, we see a square root improvement in the instance size $\lvert C_{2^n} \rvert$. Additionally, the function $f$ from the statement of Definition \ref{def:serial_adaptive} simply reads the adaptive output measurements as a binary string and returns the corresponding integer (the channel's hidden index).

			\begin{figure}[htbp!]
				\begin{center}
					\includegraphics[width=1.0\columnwidth]{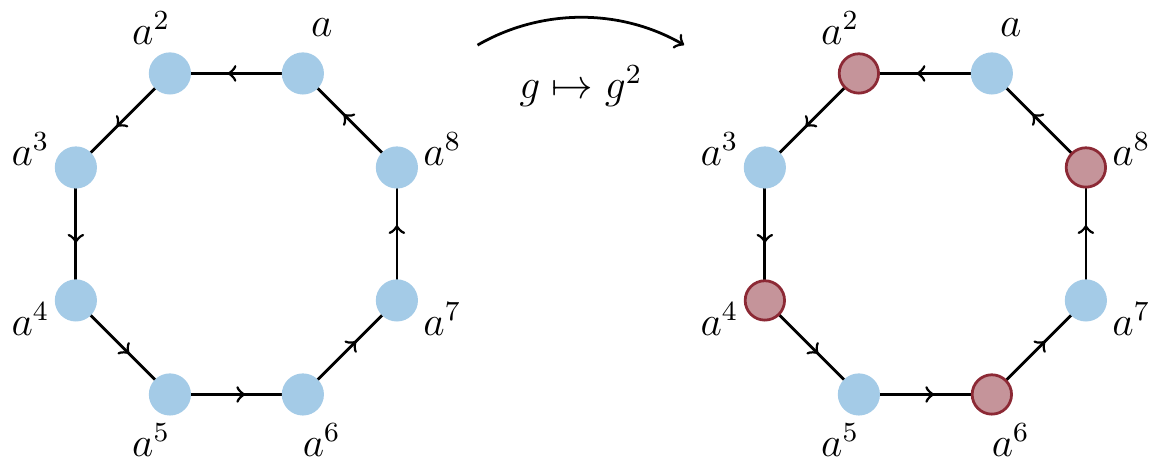}
				\end{center}
	  			\caption{Subroutine of decision protocol on $C_8$. For the cyclic group of order $2^n$, any map $g \mapsto g^{2^m}$ for $m < n$ generates a cyclic subgroup of order $2^{n - m}$. Consequently, as the cyclic group of order 2 has an obvious discrimination strategy, the method in Lemma \ref{lemma:simple_cyclic_group} can recursively determine membership of the hidden element in cosets of cyclic subgroups of $C_{2^n}$. Equivalently, the querent performs binary search, i.e., using $2^{n-1}$ queries, she can determine membership of the hidden element in the red (image) or blue (complement of the image) subset as pictured for the case $n = 3$, assuming she can solve the $n = 2$ case.}
	  			\label{figure:simple_cyclic_group}
			\end{figure}

		The reason that the simple method of Lemma \ref{lemma:simple_cyclic_group} works is because even powers of channel elements are not only subsets but subgroups of $C_{2^n}$, and specifically $2^{n-1}$ powers of group elements are rotations by angles in $\{0, \pi\}$, which give perfectly orthogonal and thus perfectly distinguishable states when acting on special known initial pure states. The adaptive protocol permits the querent to recurse and learn the hidden index by asking individual questions of coset membership for prime-power order normal subgroups.

		For cyclic groups of general order, however, this method fails. For odd-order cyclic groups, for instance, sets of integral powers of group elements do not necessarily form non-trivial subgroups by simple consequences of Lagrange's theorem. The question of bisecting the search space must thus be resolved by other methods; it is precisely the flexibility of QSP that will permit the recovery of algorithms of the same flavor as Lemma \ref{lemma:simple_cyclic_group} for more general groups. That is, to permit the construction of quantum algorithms that act deterministically on not merely subgroups but arbitrarily chosen subsets of the query set.
    
	\subsection{Paper outline and summary of results}

		The main body of this work describes methods for perfectly deciding sets of quantum channels (equivalently \emph{query sets}) which faithfully represent finite subgroups $G < \text{SU(2)}$ in order of increasing complexity of the finite group considered. This culminates in Theorem \ref{theorem:main_theorem}.
		    \begin{theorem} \label{theorem:main_theorem}
				[Simplified] There exist quantum algorithms in the serial adaptive query model which perfectly decide on all finite subgroups $G$ of SU(2), with the exception of the simple non-abelian group $A_5$, and which do so with asymptotically optimal query complexity. These algorithms each closely track with a single \emph{generic algorithm} (Algorithm \ref{algorithm:generic_algorithm}), and their individual structure closely tracks the structure of the considered group.
			\end{theorem}
		This work is organized such that algorithms for deciding simpler finite groups can, where applicable, be used as subroutines for algorithms deciding more complicated groups whose subgroup decomposition is non-trivial. It is this \emph{bootstrapped approach} that provides novel sufficient conditions under which the open question in Subsection \ref{subsection:prior_work} can be resolved in the serial adaptive query model.
		
		We begin with an overview of the two mathematical techniques that underlie the main results of the paper. Namely, in Section \ref{section:qsp_overview} we review statements of the main theorems of quantum signal processing, their guarantees, and interpretations. Relatedly, we give a protocol (Algorithm \ref{algorithm:generic_algorithm}) that players of a simplified version of the G-QHT game (Problem \ref{g_qht_problem}) defined in Subsection \ref{subsection:problem_statement} may use to achieve perfect decision protocols. The theorems of QSP (and consequently solutions to the simplified game proposed in Problem \ref{r_qht_problem}) rely on the existence and efficient computability of polynomials over real variables under simple constraints, the properties of which are discussed in Section \ref{section:poly_interpolation}.
		
		With both of the mathematical techniques established in Sections \ref{section:qsp_overview} and \ref{section:poly_interpolation}, the paper proceeds to discuss concrete groups systematically. The statement of Problem \ref{g_qht_problem} as mentioned is simplified to Problem \ref{r_qht_problem}, whose solution using the methods of QSP depends solely on the answer to questions in functional approximation. For each concrete algorithm corresponding to deciding each finite subgroup $G < \text{SU(2)}$ in Section \ref{section:explicit_construction}, we perform reductions to decisions on normal subgroups of $G$ where possible, and restate decision algorithms on $G$ as multiple correlated instances of Algorithm \ref{algorithm:generic_algorithm}. Specifically, we assert that Algorithm \ref{algorithm:generic_algorithm} and its performance guarantees are integral to the analysis of each algorithm given in Section \ref{section:explicit_construction}.
		
		Algorithm \ref{algorithm:generic_algorithm} connects decisions on $G$ to problems in functional approximation which, referring back to the guarantees of Section \ref{section:poly_interpolation}, determine the query complexity of the algorithm deciding on $G$. This connection is made explicit in Problems \ref{p_qht_problem} and \ref{extended_p_qht_problem}.
		
\begin{figure*}[htbp!]
				\begin{center}
					\vspace{1em}
			    	\includegraphics[width=0.75\textwidth]{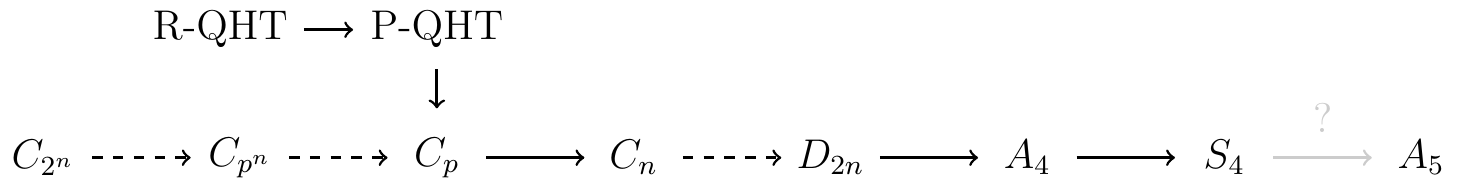}
				\end{center}
	  			\caption{The linear flow of this work: deciding on increasingly rich sets of finite subgroups of SU(2). The diagram indicates the order in which instances of G-QHT are solved throughout Section \ref{section:explicit_construction}, beginning with cyclic groups and working toward the dihedral and platonic groups; solid arrows indicate increasing complexity of the decision group, while dotted lines indicate where a reduction to an algorithm deciding on the latter group is particularly simple. R-QHT (Problem \ref{r_qht_problem}) and P-QHT (Problems \ref{p_qht_problem} and \ref{extended_p_qht_problem}) are developed in parallel to decision protocols on cyclic groups, and are joined for decisions on prime order groups by Theorem \ref{theorem:all_prime_order}. Applying similar methods to $A_5$ is left to future work.}
	  			\label{figure:problem_reduction}
			\end{figure*}

\begin{figure}[htbp!]
				\begin{center}
					\includegraphics[width=1.0\columnwidth]{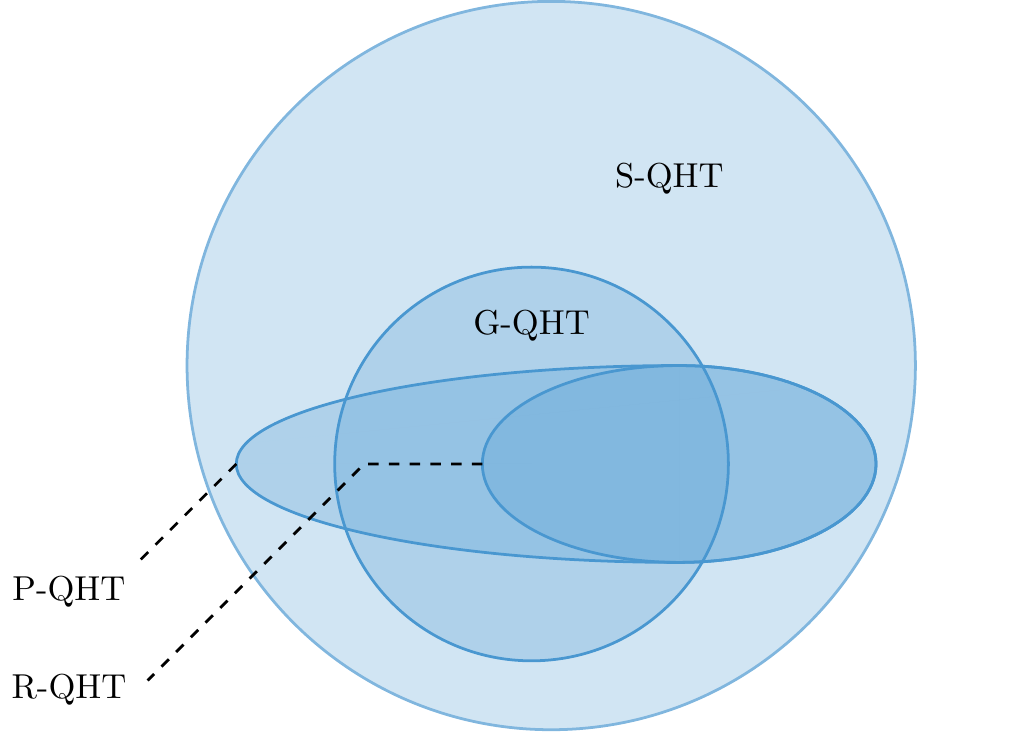}
				\end{center}
	  			\caption{Inclusion relations among problems formally defined in this work. Four major problems discussed: S-QHT (Problem \ref{s_qht_problem}), G-QHT (Problem \ref{g_qht_problem}), R-QHT (Problem \ref{r_qht_problem}), and P-QHT (Problems \ref{p_qht_problem} and \ref{extended_p_qht_problem}), referring to set, group, rotation, and polynomial quantum hypothesis testing respectively. Each region in the inclusion diagram contains non-trivial instances.}
	  			\label{figure:problem_diagram}
			\end{figure}

		We provide a diagram of the order in which we address decisions on specific finite subgroups (Figure \ref{figure:problem_reduction}) as well as relations between all problems introduced in this work (Figure \ref{figure:problem_diagram}). In turn, the relations between algorithms and problems are summarized in the statement of Algorithm \ref{algorithm:generic_algorithm} in conjunction with its accompanying remarks (Remarks \ref{remark:algorithm_remark}, \ref{remark:bisection_remark}), toward a coherent framework for hypothesis testing on discrete query sets.

		For generalizations to larger Hilbert spaces, near-unitary channels, and groups not embeddable in SU(2), the reader is directed to Section \ref{section:generalizations}. Additionally, Section \ref{section:conclusions} gives a list of open problems in the same vein as the results presented in this work, suggestions for the shape of their resolution, and instances (e.g., Remark \ref{remark:reference_remark}) in which the methods derived here can be directly applied to physical problems.

\section{Overview of quantum signal processing} \label{section:qsp_overview}

	We have defined the G-QHT problem (Problem \ref{g_qht_problem}) as well as the form that any algorithm in the serial adaptive query model solving this problem must take. We have not, however, provided a method for analyzing such algorithms. For certain groups, e.g., $C_{2^n}$ as covered in Lemma \ref{lemma:simple_cyclic_group}, we can come up with methods inspired by classical algorithms; this intuition breaks down for more complicated groups. In this section we introduce techniques toward addressing this breakdown.

	G-QHT might be naturally thought of as a sensing problem: given an unknown $g$, application of the channel $U_g$ (respecting a representation) might be physically explained as the result of probing a system: the action of the quantum channel contains some information about the system. Successive queries increase knowledge of the hidden parameter $g$ of the group action. Naturally, the ideal method for extracting information from the queried channel varies with the structure of $G$. Taking inspiration from algorithms for quantum sensing in the serial query model, we thus might naturally consider the flexible, recently developed techniques of quantum signal processing (QSP) \cite{low-16, low-chuang, low-19, gilyen}.

	QSP is a powerful quantum algorithmic primitive to implement matrix polynomials on quantum computers under only mild constraints \cite{gilyen}. Analysis of QSP has enabled intuitive constructions for asymptotically optimal algorithms in a range of settings from Hamiltonian simulation \cite{low-chuang} to the quantum linear system problem \cite{harrow-09} in \cite{gilyen, dmwl_20, lin-tong}. For our purposes, however, we will need only to consider the guarantees of the form of QSP protocols, succinctly stated in the following two theorems. Before this we briefly address an issue of notation.

		\begin{definition}
			In this work the convention when referring to the \emph{Pauli operators} is
				\begin{equation} \label{eq:pauli_operators}
					\sigma_x = 
						\begin{pmatrix}
							0 & 1 \\[-0.4em]
							1 & 0 
						\end{pmatrix}
					\quad \sigma_y = 
						\begin{pmatrix}
							0 & -i \\[-0.4em]
							i & 0 
						\end{pmatrix}
					\quad \sigma_z =
						\begin{pmatrix}
							1 & 0 \\[-0.4em]
							0 & -1 
						\end{pmatrix},
				\end{equation}
			and moreover we will often refer to a linear combination of such operators following the convention
				\begin{equation}
					\sigma_{\xi} \equiv \sigma_x\, \cos{\xi} + \sigma_y\, \sin{\xi},
				\end{equation}
			where this construction will often be used in the context of defining a rotation about a fixed axis on the Bloch sphere, namely
				\begin{equation}
					R_{\xi}(\theta) \equiv \exp\{-i(\theta/2)\sigma_\xi\},
				\end{equation}
			where this is distinct from the convention of \cite{gilyen}. If the index is Latin instead of Greek, e.g., $R_{x}(\theta)$, then it is meant $exp\{-i(\theta/2)\sigma_x\}$: rotation about the $\hat{x}$ vector.
		\end{definition}

		\begin{theorem} \label{theorem:qsp_matrix_form}
			In \cite{gilyen}. Let $k \in \mathbb{N}$; there exists $\Phi \in \mathbb{R}^{k + 1}$ such that for all $x \in [-1, 1]$
	            \begin{eqnarray}
	                e^{i\phi_0\sigma_z}\prod_{j = 1}^{k}&&\left(W(x)\,e^{i\phi_j\sigma_z}\right) =\nonumber\\
	                &&\begin{pmatrix}
	                    P(x)                    & iQ(x)\sqrt{1 - x^2}\\
	                    iQ^*(x)\sqrt{1 - x^2}   & P^*(x),
	                \end{pmatrix},
	            \end{eqnarray}
	        iff $P, Q \in \mathbb{C}[x]$ satisfy the following properties:
                \begin{enumerate}
                    \item $\text{deg}(P) = k$ and $\text{deg}(Q) = k - 1$.
                    \item $P$ has the same parity as $k$ modulo 2, while $Q$ has the opposite parity.
                    \item For all $x \in [-1, 1]$, $P$ and $Q$ satisfy $P(x)P^*(x) + (1 - x^2)Q(x)Q^*(x) = 1$.
	           \end{enumerate}
		\end{theorem}

	Theorem \ref{theorem:qsp_matrix_form} asserts that QSP protocols, which involve interleaving rotations about orthogonal axes (one of these rotations by a fixed, unknown angle, and the other by an unfixed, known angle) result in unitary operators whose elements are polynomials of the unknown rotation angle. These polynomials are under constraints necessary and sufficient to ensure the resulting operator is unitary. While the constraints of Theorem \ref{theorem:qsp_matrix_form} are non-intuitive for one wishing to solve the reverse problem (i.e., go from polynomial to a unitary operator in which the polynomial is embedded), the following theorem addresses precisely this concern.

		\begin{theorem} \label{theorem:qsp_real_polynomial_reduction}
			In \cite{gilyen}. Let $k$ in $\mathbb{Z}^+$ and let $P^\prime, Q^\prime \in \mathbb{R}[x]$; there exists some $P, Q \in \mathbb{C}[x]$ satisfying the requirements of Theorem \ref{theorem:qsp_matrix_form} such that $P^\prime = \mathfrak{R}(P), Q^\prime = \mathfrak{R}(Q)$ iff $P^\prime, Q^\prime$ satisfy the first two requirements of Theorem \ref{theorem:qsp_matrix_form} and additionally $P^\prime(x)^2 + (1 - x^2)Q^\prime(x)^2 \leq 1$.
	    	
	    	The proof of this statement follows constructively from a provably efficient (e.g., polynomial in $k$) algorithm to build the missing complex parts of $P, Q$.
		\end{theorem}
	
	In Theorem \ref{theorem:qsp_real_polynomial_reduction} the operator $W(x)$, the signal being processed, will be analogous to the quantum channel $\mathcal{E}_i$ we wish to discriminate in G-QHT. That said, the utility of these theorems is not immediately clear: the form of $W(x)$ (rotation about a known, fixed axis) is far simpler than the members of the query set considered in G-QHT for arbitrary finite subgroups of SU(2). 

	In the interest of making progress, we can thus modify the statement of Problem \ref{g_qht_problem} such that QSP stands a fair chance of providing a solution. Specifically we can write out the generic form of a QSP-based algorithm that perfectly decides any finite set $S = \{R_x(\theta_\ell)\}_{\ell \in [n]} \in [-\pi, \pi)^{n}$ under the map $R_x(\theta_\ell) = \exp\{-i\theta_\ell/2\sigma_x\}$. Note that here $S$ need not be a group under composition. This modified version of the G-QHT game is discussed in Problem \ref{r_qht_problem}.

		\begin{problem} \label{r_qht_problem}
			The rotation QHT problem (\emph{R-QHT problem}) is a simplified version of the G-QHT problem (Problem \ref{g_qht_problem}) with the following structure. Given query access to a single-qubit quantum channel from among a finite set $S$ where each channel has again the form $R_{\xi}(\theta_i) = \exp\{-i(\theta_i/2)(\cos{\xi}\sigma_x + \sin{\xi}\sigma_y))\}$ for distinct, known $\theta_i$ and known rotation axis $\xi$, determine the queried channel with certainty in the serial adaptive query model.

			Note that R-QHT problems are not a subset of G-QHT problems, save in the case that the set of angles $\{\theta_\ell\}$ are all distinct integral multiples of $2\pi/n$ for positive integral $n$ (i.e., $S$ represents a cyclic group).
		\end{problem}

	As the rotation operators discussed in the R-QHT problem satisfy the form expected of the $W(x)$ operator in QSP, the methods of QSP suggest a neat prescription for a quantum algorithm (Algorithm \ref{algorithm:generic_algorithm}) with classical subroutines such that the output is a solution for the R-QHT problem. We discuss assumptions on the input, output, and structure of Algorithm \ref{algorithm:generic_algorithm} in Remark \ref{remark:algorithm_remark}, give definitions for its classical subroutines in Definition \ref{def:r_qht_algorithm}, and further remark on where the non-trivial aspects of Algorithm \ref{algorithm:generic_algorithm} lie in Remark \ref{remark:bisection_remark}.

		\begin{remark} \label{remark:algorithm_remark}
			
			We present a series of data structures which together define both an instance of the R-QHT problem (Problem \ref{r_qht_problem}) and its solution, toward a concrete algorithm (Algorithm \ref{algorithm:generic_algorithm}).

			\begin{itemize}

				\item \textbf{Input}: Any instance of R-QHT presupposes access to \emph{classical information} in the form of a list of distinct angles $\{\theta_\ell \in [0, 2\pi]\}, \ell \in [n]$. R-QHT also presupposes access to a \emph{quantum oracle} which, when called, applies a quantum channel channel $R_{\xi}(\theta_i)$ for fixed $i$ about some known fixed axis $\xi$.

				\item \textbf{Output}: In the serial adaptive query model on qubits, a projective measurement is an evaluation of a probabilistic binary function on possible hidden indices $j \in [n]$ for the applied channel. An R-QHT algorithm's output is one of these indices, where success is dictated by high probability\footnote{In the noiseless case, we consider only deterministic algorithms.} of or certainty in returning the proper hidden index $i$.

				\item \textbf{Assumptions}: The result of the evaluation of a set of these functions (corresponding to $m$ binary measurements), $f_j : [n] \mapsto \{0, 1\},\; j \in [m]$ on the hidden index $i$ of the queried channel, is a composite function $g: i \mapsto \{0, 1\}^{m}$ defined as $g(i) = f_1(i)f_2(i)\cdots f_m(i)$. 

				If this function is injective for all $j \in [n]$ then the algorithm generating the $f_j$ solves R-QHT.\footnote{This is a non-trivial condition to satisfy, but in most instances can be thought of as assigning a binary tree's labels to each of $m$ channels. This is the subject of Remark \ref{remark:bisection_remark}.} Equivalently the algorithm computes a series of $m$ equivalence relations on the set of rotation angles $\{\theta_\ell\}, \ell \in [n]$ such that every element is uniquely defined by its membership under these $m$ bisections.

			\end{itemize}

		\end{remark}

		\begin{definition} \label{def:r_qht_algorithm}
			A quantum algorithm solving the R-QHT problem (Problem \ref{r_qht_problem}) is referred to simply as an \emph{R-QHT algorithm}, where \emph{solves} indicates that it satisfies the input, output, and structural assumptions presented in Remark \ref{remark:algorithm_remark}.

			In addition, toward an explicit description of one such \emph{R-QHT algorithm} (Algorithm \ref{algorithm:generic_algorithm}), we define four classical sub-algorithms whose application together constitutes the classical subroutine of Algorithm \ref{algorithm:generic_algorithm}).
				
				\begin{itemize}
					
					\item \emph{\texttt{genBisection}}: Given a group representation $G$ and a (possibly empty) set of evaluations of previous binary functions $f_j: S_j \rightarrow \{0, 1\}$ for $S_j \subseteq S_{j - 1} \subseteq \cdots \subseteq S_1 \subseteq G$, returns \emph{a description} of $f_{j + 1}: S_{j + 1} \rightarrow \{0, 1\}$ where $S_{j + 1} \subseteq S_{j}$ is a subset of $S_{j}$ on which $f_{j}$ is constant.

					The choice of $f_{j + 1}$ is not arbitrary but instead depends heavily on the embedding of $G$ in a larger continuous group. Examples for methods of choosing these $f_j$ can be found in the concrete algorithms of Section \ref{section:explicit_construction}. Further discussion of the properties of these functions is also covered in Remark \ref{remark:bisection_remark}.
					
					Note that in Algorithm \ref{algorithm:generic_algorithm}, the description of $f_{j + 1}$ can be used to compute $f_{j + 1}(i)$ on the hidden index, oblivious to the hidden index.
					
					\item \emph{\texttt{genRealPoly}}: Given a description of $f_j$, defined on some subset of group elements $S_j \in G$, where each $s \in S_j$ is parameterized by some distinct real parameter $\theta_\ell \in [0, 2\pi]$ for $\ell \in \lvert S_j\rvert$, returns the minimal degree real polynomial $p_j$ satisfying $\lvert p_j(\arccos\theta_\ell)\rvert = f_j(s[\theta_{\ell}])$ for all $\theta_\ell$, and where $\lvert p_j(\theta) \rvert \leq 1$ for $\theta \in [0, 2\pi]$. In addition $p_j$ is of definite parity on $[-1,1]$.

					Methods for computing constrained interpolating polynomials are numerous and well-studied, comprising the discussion of Section \ref{section:poly_interpolation}.

					\item \emph{\texttt{genComplexPoly}}: Given a real polynomial $p_j$ satisfying the constraints of the output of \emph{\texttt{genRealPoly}}, returns a pair of complex polynomials $(P_j, Q_j)$ on $[-1,1]$, each of definite parity and satisfying $P_j(x)^2 + (1 - x^2)Q_j(x)^2 = 1$ for $x \in [-1, 1]$. Moreover $\mathfrak{R}(P_j) = P^\prime_j = p_j$ and $\mathfrak{R}(Q_j) = 0$. One implementation is given in \cite{gilyen}.

					\item \emph{\texttt{genPhases}}: Given two polynomials $(P_j, Q_j)$ satisfying the constraints on the output of \emph{\texttt{genComplexPoly}}, returns a set of phase angles $\Phi_j \in \mathbb{R}^{k + 1}$ satisfying Theorem \ref{theorem:qsp_matrix_form}.
					
					This subroutine also returns a \emph{classical description} of two quantum states, $\psi_j, \psi_{j}^\prime$, the former an initial state and the latter a state with respect to which a projective measurement is performed to compute $f_j$ on the hidden index, i.e., $f_j(i)$. These states are efficiently computable and project out $p_j$, equivalently $\braket{\psi_{j}^\prime \lvert U_{\Phi_j} \rvert \psi_j} = p_j$, where $U_{\Phi_j}$ is the QSP unitary generated by $\Phi_j$.

					Methods for computing these phase factors are numerous \cite{gilyen, dmwl_20, haah_19}; all affirm that this computation is efficient and stable, using existing techniques in classical optimization.

					\item We denote by $M_{\psi_j}(\ket{\psi})$ the measurement projecting $\ket{\psi}$ onto $\{M_0, M_1\} = \{\lvert \psi_j \rangle\langle \psi_j \rvert, I - \lvert \psi_j \rangle\langle \psi_j \rvert\}$, returning $b$ upon measurement of $M_b$.
				\end{itemize}
				
		\end{definition}
        
		\begin{algorithm}
			\DontPrintSemicolon
			\BlankLine
			\SetKwInOut{Input}{Input }
			\SetKwInOut{Output}{Output }
			\SetKwInOut{Assumptions}{Assumptions }
				\Assumptions{Input and output satisfying assumptions of Remark \ref{remark:algorithm_remark}
				}
				\Input{A quantum channel oracle $\mathcal{E}_i$ for hidden index $i$; description of $n$ channels $\{\mathcal{E}_\ell\}_{\ell \in [n]}$.
				}
				\Output{The hidden channel index $i$.
				}
			\BlankLine
			\For{$j\leftarrow 1$ \KwTo $m$}{
				\BlankLine
				\textbf{Classical subroutine (see Def.\ \ref{def:r_qht_algorithm}):}
				\BlankLine
					$f_j \leftarrow \texttt{genBisection}(G, \{f_{ < j}(i)\})$\;
				\BlankLine
					$p_j \leftarrow \texttt{genRealPoly}(f_j)$\; 
				\BlankLine
					$(P_j, Q_j) \leftarrow \texttt{genComplexPoly}(p_j)$\; 
				\BlankLine
					$(\Phi_j, \psi_{j}, \psi_{j}^\prime) \leftarrow \texttt{genPhases}(P_j, Q_j)$\;
				\BlankLine
				\textbf{Quantum subroutine:}
				\BlankLine
					$\ket{\psi} \leftarrow \ket{\psi_{j}}$
					\emph{Initialize quantum state}\;
					\BlankLine
					\For{$k\leftarrow 1$ \KwTo $n_j$}{
							$\ket{\psi} \leftarrow R_{\xi}(\theta_i)\ket{\psi}$
							\emph{Apply oracle for unknown $i$}\;
							$\ket{\psi} \leftarrow U_k\ket{\psi}$
							\emph{Apply QSP unitary $ \exp\{i\phi_k\sigma_{\xi^\perp}\}$}\;
						}
					\BlankLine
					$f_j(i) \leftarrow  M_{\psi_{j}^{\prime}}(\ket{\psi}) \;\; \text{\emph{Send} $\{\ket{\psi_j^\prime}, \ket{\psi_j^{\prime\perp}}\} \mapsto \{0, 1\}$ }$ \;
				\BlankLine
			}
			\BlankLine
			$i \leftarrow g(i) = f_1(i)f_2(i)\cdots f_m(i)$ \emph{Invert $g$ by Remark \ref{remark:algorithm_remark}}\;
			\BlankLine
			Return $i$\;
			\caption{A generic algorithm for solving R-QHT\vspace{0.5em}}
			\label{algorithm:generic_algorithm}
		\end{algorithm}
			
		\begin{remark} \label{remark:bisection_remark}
			The difficulty in Algorithm \ref{algorithm:generic_algorithm} stems from selection of the proper functions $f_j: S_j \rightarrow \{0, 1\}$ for subsets $S_j \subseteq S$ of the query set of fixed-axis rotations (equivalently computing \emph{\texttt{genBisection}} in Definition \ref{def:r_qht_algorithm}).

			As each $f_j$ takes values on $S_j$ in $\{0, 1\}$, they can be thought of as labels dividing or bisecting the query set; the result of QSP is to make the quantum computation of these $f_j$ on the hidden index $i$ deterministic. A series of these $f_j$ thus form the levels of a binary tree whose bisection condition is the result of a projective measurement onto $\{\ket{\psi_j^\prime}, \ket{\psi^{\prime\perp}_j}\}$. We discuss the desired properties of this binary decision tree; these principles foreshadow the properties discussed in Theorem \ref{theorem:all_prime_order}.

				\begin{itemize}
					
					\item An efficiently searchable binary tree should be balanced; different channels should have binary labels according to the tree which differ as early as possible, equivalently each $f_j$ should divide the remaining query set roughly in half.
					
					\item The discrete $f_j$ objects are accessed by interpolating polynomials in a \emph{continuous embedding space}, and as the minimal degree of such polynomials correspond to algorithmic performance, we desire that the $f_j$ subdivide the search space into subsets which have a larger average\footnote{This is purposefully left ambiguous at this moment; we wish to lower the required derivative of the interpolating polynomial.} distance between elements in the natural metric of this space. Equivalently proximate elements in the binary tree are also proximate in the embedding space.
					
					\item Each leaf of the binary decision tree must correspond to no more than one channel. If each (probable with respect to measurement) leaf corresponds exactly to one channel, then $g$ in \ref{remark:algorithm_remark} is not only injective but bijective.
					
					\item The $f_j$ must have definite parity in the continuous embedding space, here SU(2); this parity constraint, requisite for the use of QSP, follows from properties of SU(2).

				\end{itemize}

		\end{remark}

	Algorithm \ref{algorithm:generic_algorithm} and its supporting remarks show that, at least for a special set of channels, our hopes of computing successive equivalence relations on subsets of $S$ to iteratively determine the hidden query element rest on the construction of low-degree constrained polynomials over real variables.

	Moreover, as stated in Remark \ref{remark:bisection_remark}, most of the difficulty of this algorithm resides in designing the binary functions $f_j$. The sequence of equivalence relations $f_1, f_2, \cdots , f_m$, which together uniquely define the hidden index $i$, need to be properly chosen such that (1) the degrees of their polynomial interpolations are not too large, and (2) that the concatenation of their evaluations is invertible on every $i$; luckily these conditions are not so complicated to achieve in practice.

	E.g., we can see one such set of $f_j$ in observing the `QSP-free' decision algorithm for $C_{2^n}$ in Lemma \ref{lemma:simple_cyclic_group}, namely $f_j(i) = i \pmod{2^{j}}$ for $j \in [n]$. Evidently in this simplest case the family of $f_j$ define precisely a binary search on the hidden channel index (and consequently the equator of the Bloch sphere under the map $i \mapsto \mathcal{E}_i$). What remains to be shown is the generalization of such a search.

		\begin{figure*}[htbp!]
			\begin{center}
		    	\includegraphics[width=0.75\textwidth]{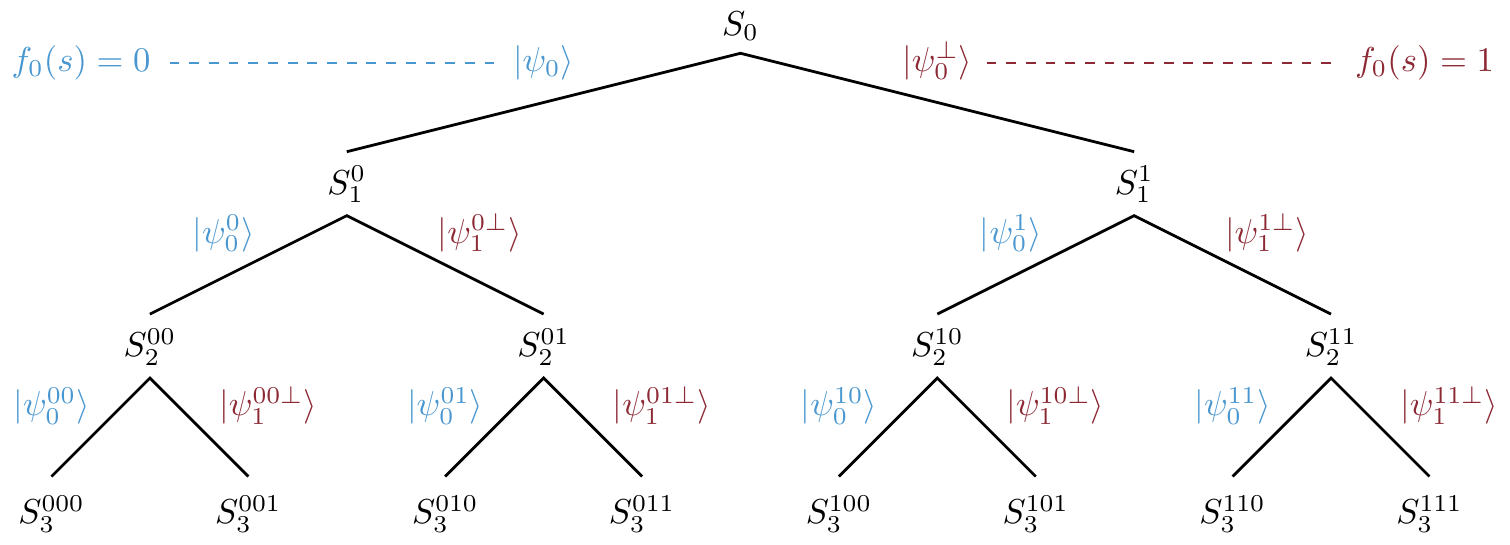}
			\end{center}
			\caption{Binary search as enacted by Algorithm \ref{algorithm:generic_algorithm}. Algorithm \ref{algorithm:generic_algorithm} takes binary functions $f_j$ on subsets of $S_0$, specifically $S_j^{f_{< j}(s)}$, and produces a quantum algorithm that maps elements $s$ on which $f_j$ takes value $\{0, 1\}$ to orthogonal quantum states $\{\ket{\psi_j^{f_{< j}(s)}}, \ket{\psi_j^{f_{< j}(s)\perp}}\}$ respectively. Measurement in this basis determines the new query set, $S_{j + 1}^{f_{< j + 1}(s)}$, and the process is repeated until each leaf of the binary tree contains at most one quantum channel. The notation $s$ here is overloaded, indicating both the quantum channel and the continuous parameter defining the channel. For extension of this concept from R-QHT to G-QHT see Remark \ref{remark:recursion_remark}}
			\label{fig:decision_tree}
		\end{figure*}

	It turns out that Algorithm \ref{algorithm:generic_algorithm} can indeed be extended to more interesting channel sets than single-axis rotations (i.e., that we can lift R-QHT problems to G-QHT problems). However, before investigating the flexibility of Algorithm \ref{algorithm:generic_algorithm} as a subroutine, we first briefly address methods in constrained polynomial interpolation. This analysis, in addition to closing the loop on the R-QHT problem and its query complexity, will demonstrate the methods by which the optimal query complexity of R-QHT is computed, and provide a foundation for generalizing to G-QHT.

\section{Constrained polynomial interpolation} \label{section:poly_interpolation}

	In the previous section we reduced the solution of Problem \ref{r_qht_problem}, a simplified version of G-QHT, to the existence of interpolating polynomials over real intervals. Moreover we asserted that, despite the restrictive form of the queried channel $W(x)$ considered in QSP, the guarantees of Theorem \ref{theorem:qsp_matrix_form} were still strong enough to enable discrimination among channel sets whose structure is richer than rotations about a fixed axis. This section considers one concrete interest of a party playing R-QHT: how can a computationally limited classical party compute $\Phi$ for a QSP algorithm such that the resulting matrix polynomials induce measurements obeying the prescriptions of Algorithm \ref{algorithm:generic_algorithm}.

	This is a problem of constrained polynomial interpolation. More generally, the field of functional approximation, in which this problem lives, is well-understood \cite{erdos, wolibner, mhaskar-sain, mclaughlin-sain, deutsch-sain, yamabe} given its practical instantiations in classical signal processing and relevance to foundational questions in real analysis. We quote the following results in constrained polynomial approximation and present their synthesis as a new theorem guaranteeing desired properties for the algorithms that will be constructed in Section \ref{section:explicit_construction} for specific finite groups. Additionally, these results provide quantitative bounds on the query complexity of solutions to the R-QHT problem discussed previously.

	We present a further sharpening of R-QHT (Problem \ref{r_qht_problem}); this new problem, P-QHT, is similar to R-QHT but provides a new quantitative condition on the performance of an algorithm solving R-QHT.
    	\begin{problem} \label{p_qht_problem}
    		The polynomial QHT problem, or \emph{P-QHT} problem, answers the following question. Given an instance of the R-QHT problem (Problem \ref{r_qht_problem}), which implicitly defines a set of angles $\{\theta_\ell\}$, what is an upper bound on the sum of degrees of the set of polynomials $\{p_j\}$ which interpolate binary functions\footnote{Note that for our purposes it is often not important to distinguish between $\{\ell\}$ the set of indices and $\{\theta_\ell\}$ the set of angles. While the degree of the interpolating polynomial depends on these angles, this dependence can be simplified by promises on separations between neighboring $\theta_\ell$.} $f_1, f_2, \cdots, f_m$ satisfy Remark \ref{remark:bisection_remark}. This upper bound depends only on $\{\theta_\ell\}$.
    	\end{problem}
	Toward analyzing the minimal degree of such interpolating polynomials as desired in Problem \ref{p_qht_problem}, we give a series of older results from works in constrained interpolation.
		\begin{theorem} \label{theorem:polynomial_existence}
			\emph{In \cite{wolibner}} Let $\Xi = \{x_i \,:\, x_1 < x_2 <  \cdots < x_n\}$ a set from the real interval $[a, b]$ and $\mathcal{P}$ the set of polynomials. For all $\epsilon > 0$ and for each $f \in C^0[a, b]$, the continuous functions on $[a, b]$, there exists $p \in \mathcal{P}$ such that the following conditions are satisfied:
			    \begin{enumerate}
				    \item $p$ is interpolating: $p(x_i) = f(x_i)\; \forall x_i$.
				    \item The polynomial $p$ uniformly approximates $f$ on $[a, b]$,
    					\[
    						\max_{x \in [a, b]}\; \lvert p(x) - f(x) \rvert \leq \epsilon.
    					\]
				    \item The polynomial $p$ obeys the additional constraint
    					\[
    						\max_{x \in [a, b]}\; \lvert p(x) \rvert = \max_{x \in [a, b]}\; \lvert f(x) \rvert.
    					\]
				\end{enumerate}
		\end{theorem}

		\begin{theorem} \label{theorem:polynomial_order}
			\emph{In \cite{beatson}} Let $\nu$ index an increasing sequence of finite dimensional approximation subspaces $N_\nu$ in $C(T)$, for $T$ a topological space, whose union $N$ is dense in $C(T)$. If $T$ is compact Hausdorff then the degree of approximation with Lagrange (function value) interpolatory side conditions $E_\nu(f, A)$ is related to the degree of the unrestricted approximation $E_\nu(f)$ by the inequality
				\[
					\limsup_{\nu\rightarrow\infty} \frac{E_\nu(f, A)}{E_\nu(f)} \leq 2 \quad \forall f \in C(T)\backslash N,
				\]
			where the constant $2$ cannot be decreased in general, and is the best possible in the uniform approximation of (1) entire periodic functions by trigonometric polynomials and (2) entire functions on any closed finite interval by algebraic polynomials.
		\end{theorem}

		\begin{corollary} \label{beatson_corollary}
			In the context of constrained polynomial interpolation the statement of Theorem \ref{theorem:polynomial_order} can be made less general as follows: Given a real interval $[a, b]$ and a real polynomial $f$ of degree $d$ which interpolates a function $g$ on $[a, b]$ at $d$ distinct points in $[a, b]$, the minimal degree of a polynomial which interpolates $g$ at these same points and has norm strictly less than $\lVert g \rVert$ on $[a,b]$ is bounded above by $2d$ as $d$ goes to infinity and moreover this bound cannot be decreased in general.
		\end{corollary}

		\begin{theorem} \label{theorem:szabados_promised_gap_interpolation}
			\emph{In \cite{mhaskar-sain}} Let $n \in \mathbb{Z}^+$ and let $x_{j} = \cos{\theta_j}$ where $\theta_{1} < \theta_{2} < \cdots < \theta_{n} \in [0, 2\pi]$ and the minimum separation between adjacent $\theta_j$ (on the unit circle) is given by $\delta > 0$.  Given any real function $f \in C([-1,1])$ there exists a polynomial $p$ such that the following conditions hold:
			    \begin{enumerate}
                    \item $p$ is interpolating: $p(x_j) = f(x_j)\; \forall x_j$.
                    \item The polynomial $p$ is of degree $2m \leq c/\delta$ where $c > 0$ is some absolute constant.
                    \item The following inequality holds where the infimum is taken over the space of all polynomials $q$ of degree at most $2m$ and $k$ is a constant independent of $f$ and $n$:
                    	\[
                    		\max_{x \in [-1,1]} \rvert f(x) - p(x) \lvert 
                    		\;\;\,\leq\; k \inf_{q \in \mathcal{P}} \left(\max_{x \in [-1,1]} \lvert f(x) - q(x) \rvert\right)
                    	\]
                \end{enumerate}
		\end{theorem}

		\begin{theorem} \label{theorem:promised_gap_interpolation}
			Let $\Xi = \{x_j\}_{j \in [n]}$ where $x_{j} = \cos{\theta_j}$ and where $\theta_{1} < \theta_{2} < \cdots < \theta_{n} \in [0, 2\pi]$ such that the minimum separation between adjacent $\theta_j$ (on the unit circle) is given by $\delta > 0$. Then given any real function $f \in C([-1,1])$ there exists a polynomial $p$ such that the following conditions hold:
			    \begin{enumerate}
				    \item $p$ is interpolating: i.e., $p(x_j) = f(x_j)\; \forall x_j$.
				    \item The polynomial $p$ is of degree $m = \mathcal{O}(1/\delta)$.
				    \item The polynomial $p$ satisfies the following inequality
    					\[
    						\max_{x \in [-1, 1]} \lvert p(x) \rvert = \max_{x \in [-1, 1]} \lvert f(x) \rvert.
    					\]
				\end{enumerate}

			\begin{proof}
				The existence of this polynomial is assured by Theorem \ref{theorem:polynomial_existence}, the scaling of degree of the unconstrained (uniformly approximating) polynomial is given by Theorem \ref{theorem:promised_gap_interpolation}, and that the constrained polynomial's degree does not grow too large with respect to the unconstrained polynomial's is given by Theorem \ref{theorem:polynomial_order}.
			\end{proof}
		\end{theorem}
	
	Finally, we present a lemma which permits us to apply all of the above results in the context, mandated by QSP, that the constrained interpolating polynomials used have definite parity.

		\begin{lemma} \label{lemma:parity_preservation}
			If there exists a polynomial of degree $n$ interpolating a set of points which has (the point set) definite parity, and the polynomial is of fixed norm, then there exists a polynomial of degree $m \leq n$ which still interpolates the points and which has the same parity as the points. Proof follows by re-expressing the polynomial as a sum of terms with definite parity and observing that the component of parity matching those of the interpolation points still satisfies the desired properties.
		\end{lemma}

	The results of this series of theorems, and particularly the assurances of Theorem \ref{theorem:promised_gap_interpolation}, permit us to justify the idealized claims of the classical program discussed in Algorithm \ref{algorithm:generic_algorithm}, at least for cyclic groups. I.e., given that the quantum channels considered can be (at least for the case that $G$ is cyclic) distinguished by their eigenvalues, the methods of QSP and the assurances of Theorem \ref{theorem:promised_gap_interpolation} together imply that their exist computationally cheap, flexible quantum algorithms whose measurement results are themselves deterministic functions on the discrete set of possible channels.

	With respect to a resolution of Problem \ref{p_qht_problem}, this section has provided a key observation: the minimal degree of the interpolating polynomial on a set of angles $\{\theta_\ell\}$, as in the R-QHT, problem is linear in both the number of interpolation points and $\max_{\ell, k} 1/\lvert \theta_\ell - \theta_k\rvert$, the minimal separation between (distinct) queried angles.

	Once the interpolating polynomials $p_j$ are computed, the path to generating QSP angles $\Phi_j$ is well understood and computationally efficient (i.e., polynomial in the degree of the interpolating polynomial). There are many ways to perform such a computation, both analytically \cite{gilyen} and by numerically stable computations \cite{dmwl_20}. Moreover, the interpolating polynomials can be computed in any number of ways, usually relating to a modified Remez-type algorithm \cite{remez, grenez}.

\section{Decision protocols on finite subgroups of SU(2)} \label{section:explicit_construction}
	
	We now close the loop on our simplification of G-QHT in Problem \ref{g_qht_problem} to R-QHT in Problem \ref{r_qht_problem} and finally, through Algorithm \ref{algorithm:generic_algorithm} to a problem in polynomial interpolation where the degree of these polynomials relates directly (by the results of Problem \ref{p_qht_problem}) to the query complexity of the solution to R-QHT. In this section we finally address the more general problem of G-QHT for small groups $G$.

	For each finite subgroup $G < \text{SU(2)}$, we provide constructive proof that there exists a \emph{series of binary functions} $f_1, f_2, \cdots, f_m$ and a \emph{series of protocols to access sets of rotations about known, fixed axes} for which the polynomials that interpolate each $f_j$ can be explicitly described, computed, and characterized in terms of degree. Once this degree is known, the expected query complexity of these algorithms follows by the results of Section \ref{section:poly_interpolation}. Before this, however, we extend the statement of P-QHT (Problem \ref{p_qht_problem}), which as stated applied only to sets of rotations about a fixed axis, to sets which obey more general structure.

		\begin{problem} \label{extended_p_qht_problem}
			The P-QHT problem (Problem \ref{p_qht_problem}) can be extended given the following prescription on a solution form. We begin with the standard statement of G-QHT: given query access to one quantum channel among a faithful representation of a finite group $G < \text{SU(2)}$ determine the optimal query complexity of an adaptive serial query model algorithm that determines the hidden index of the queried channel with certainty.

			Importantly, however, for P-QHT to provide a solution, one must be able to transform the query set in a special way; this reduction follows from the conditions given below:
				\begin{enumerate}

					\item There must exist a series of protocols, given query access to a channel set $S$, for generating \emph{compound queries}\footnote{In simple terms one may think of these as small quantum circuits which employ a small number of queries to the original oracle, and may be used as subroutines replacing oracle calls for a protocol expecting queries of a different form. Multiple \emph{physical} queries can form one \emph{compound} query.} (see Definition \ref{def:compound_queries}) whose structure is (1) precisely a set of rotations by known angles around a fixed axis (i.e., inputs to the R-QHT problem satisfying Remark \ref{remark:algorithm_remark}), or (2) a subset of a finite group $G^\prime$ for which a decision algorithm is already known.

					\item In the case of (1) as given above there must exist a solution for P-QHT (Problem \ref{p_qht_problem}) for the new query set. There must also exist some additional assumption, specific to the structure of $S$, that permits the compound query map used to be invertible. This is accomplished in different ways for different groups, e.g., under the assumption that the represented group is a semi-direct product, as in Theorem \ref{theorem:dihedral_decision}.

				\end{enumerate}
		\end{problem}
		
		\begin{definition} \label{def:compound_queries}
            A compound query with respect to a quantum channel $\mathcal{E}: A \rightarrow B$ is a quantum circuit $\mathcal{C}: A \rightarrow B$ which uses a finite number of copies of $\mathcal{E}$ as well as a finite number of additional unitary operators independent of $\mathcal{E}$.
            
            Compound queries are often used by quantum algorithms (e.g., Algorithm \ref{algorithm:generic_algorithm}) in place of bare queries, i.e., simply $\mathcal{E}_i$. Usually, useful compound query circuits do not act injectively on the query set.
		\end{definition}

		\begin{remark} \label{remark:reduction_remark}
			The extended statement of the P-QHT problem (Problem \ref{extended_p_qht_problem}) exists to answer the following question: how far can Algorithm \ref{algorithm:generic_algorithm} be taken beyond its role as a solution to R-QHT?

			Consequently each of the algorithms discussed in this section is, in truth, simply (1) a procedure for reduction to R-QHT, followed by (2) application of Algorithm \ref{algorithm:generic_algorithm}. When reduction is made to deciding a simpler group, the application of Algorithm \ref{algorithm:generic_algorithm} is hidden behind algebraic abstraction.
		\end{remark}

	We go through the finite list of distinct families of finite subgroups of SU(2) in order of increasing complexity, recovering instances of Problem \ref{extended_p_qht_problem} as stated above. As a road-map we provide the following lemma, which completely characterizes the finite subgroups of SU(2). A diagram of the path of these reductions was given in Figure \ref{figure:problem_reduction}.

		\begin{lemma} \label{lemma:finite_subgroups_su2}
			The finite subgroups of SU(2) are in bijection with the finite subgroups of SO(3) under the standard double covering $\text{SU(2)} \rightarrow \text{SO(3)}$. These finite subgroups are thus completely described by five families:
			(1) The cyclic groups of order $n$, $C_n$, $n \in \mathbb{Z}^{+}$.
			(2) The dihedral groups of order $2n$, $D_{2n}$, $n \in \mathbb{Z}^{+}$.
			(3) The alternating group on four elements, $A_4$.
			(4) The symmetric group on four elements, $S_4$.
			(5) The alternating group on five elements $A_5$.
		\end{lemma}

	\subsection{Cyclic groups}

		Before lifting the methods of Lemma \ref{lemma:simple_cyclic_group} from $C_{2^n}$ to general cyclic groups we provide a few lemmas.
			\begin{lemma} \label{cyclic_group_decomposition}
				The cyclic group of order $n$ is isomorphic to the direct product of cyclic groups
		            \[
		                C_n \cong C_{p_1^{r_1}}\times C_{p_2^{r_2}} \times \cdots \times C_{p_s^{r_s}},
		            \]
		        iff the unique prime decomposition of $n$ is
		            \begin{equation} \label{eq:prime_decomposition}
		                n = \prod_{i = 1}^{s} p_i^{r_i},
		            \end{equation}
		        for distinct primes $p_i$. I.e., $C_n$ is isomorphic to a direct product of cyclic groups of prime-power order for all maximal prime powers dividing $n$. This is one statement of the Chinese remainder theorem.
			\end{lemma}
		We proceed to analyze decisions on $C_n$ by a series of reductions to decisions on the more restricted (albeit infinite) family of cyclic groups of prime order.
			\begin{lemma} \label{lemma:prime_reduction}
				If there exists a family of algorithms $\mathcal{F} = \{\mathcal{A}_{C_p}\}$ that each perfectly decide $C_p$ for all primes $p$ then there exists an algorithm $\mathcal{A}_{C_n}$ that perfectly decides $C_n$ for $n \in \mathbb{N}$, and which is asymptotically optimal in query complexity if the algorithms in $\mathcal{F}$ are also optimal.

				\begin{proof}
					Any positive integer $n$ has a unique decomposition into a product of unique primes as given in (\ref{eq:prime_decomposition}), where $r_{i}$ is the multiplicity of the $i$-th smallest prime dividing $n$, $p_{i}$, and $s$ is the largest index for which $p_i$ divides $n$ at least once.

					Assuming the existence of a deterministic algorithm $\mathcal{A}_{C_{p_i}}$ that can perfectly decide $C_{p_i}$, elements of the group $C_n$ are decided according to the following protocol:
						\begin{enumerate}
							\item If the multiplicity $r_i$ of $p_i$ in $n$ is one, in the place of the query usually made by the protocol $\mathcal{A}_{C_{p_i}}$, query the oracle $n/p_i$ times. This compound query may be conjugated by a known unitary to achieve the representation that $\mathcal{A}_{p_i}$ expects.

							\item If the multiplicity of $p_i$ in $n$ ($r_i$) is greater than one, the same method presented in the Lemma \ref{lemma:simple_cyclic_group} is applied to compound queries of order $n/p_i^{r_i}$ to read off successive bits (this time in base $p_i$) of $r_i$, using the assumed subroutine for deciding $\mathcal{A}_{C_{p_i}}$.
						\end{enumerate}
					Compound queries allow access to prime-power-order cyclic subgroups of $C_n$ whose decision algorithms are strictly simpler and reducible to decisions on $C_p$ for $p$ prime.
				\end{proof}
			\end{lemma}
		We proceed by considering a result concerning the smallest non-trivial cyclic group, $C_3$, with which to play G-QHT. This can be thought of as a base case for our eventual reduction from decision protocols on large cyclic groups to smaller ones.
		
		Discriminating between quantum channels representing $C_3$ has some precedent in prior work: such channels are precisely those which can generate the \emph{Peres-Wootters states} \cite{shor_04, peres_wootters} or equivalently \emph{Mercedes-Benz frames} \cite{PJ_17, MN_12} (for their threefold symmetry).
		
\begin{figure}[htpb!]
				\centering
					\subfloat[Phase I]{\includegraphics[width=0.9\columnwidth]{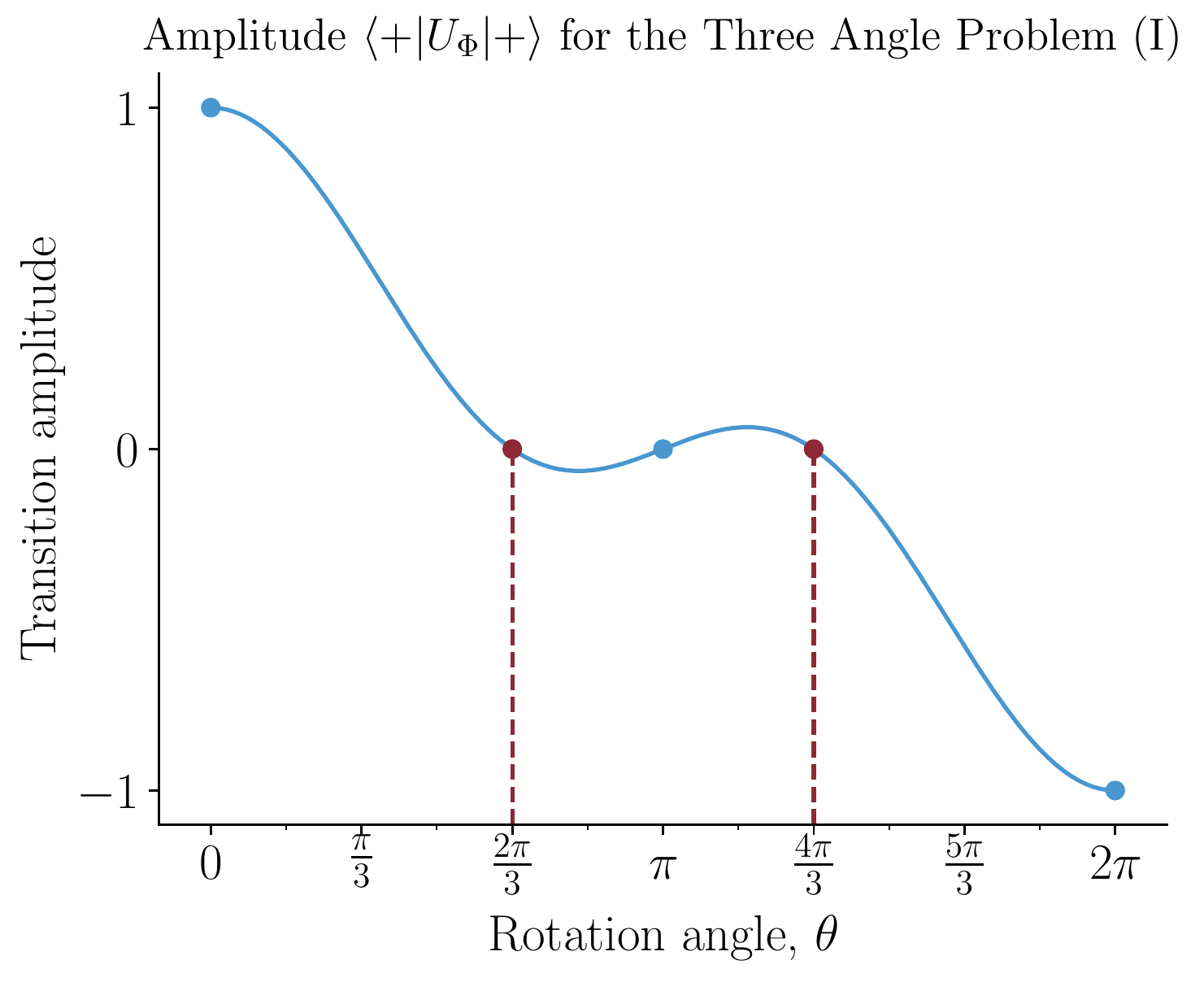}\label{subfig:three_angle}}
				\hspace{1em}
					\subfloat[Phase II]{\includegraphics[width=0.9\columnwidth]{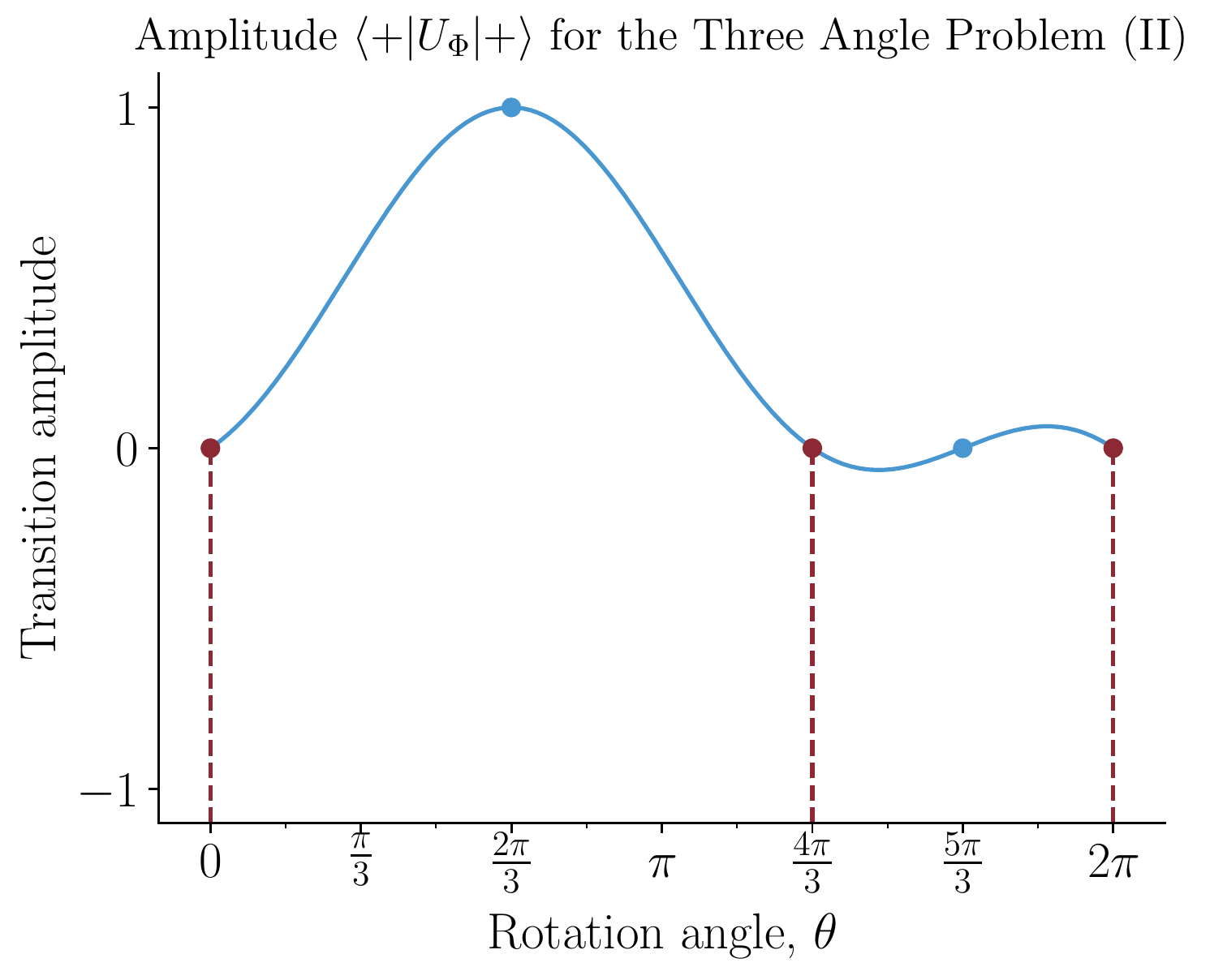}\label{subfig:three_angle_shifted}}
				\caption{Quantum response function employed in the proof of Lemma \ref{lemma:order_3} (a), and its shifted version (b). On the left is the polynomial, in $\cos{(\theta/2)}$, which is generated as the top left component of the single-qubit unitary $U_{\Phi}$ corresponding to the angles $\Phi$ indicated in the first QSP subroutine of Lemma \ref{lemma:order_3}. On the right is the same protocol using a pre-rotation by $2\pi/3$, permitting a unique binary labeling of each channel after two measurements.}
	  			\label{figure:three_angle_problem}
			\end{figure}

			\begin{lemma} \label{lemma:order_3}
				There exists an algorithm $\mathcal{A}_{C_3}$ that perfectly decides $C_3$ (or rotations about a fixed axis on the Bloch sphere by one angle among the three angle set $\{0, 2\pi/3, 4\pi/3\}$) using at most 6 oracle queries. This algorithm is said to solve the \emph{three angle problem}.

				\begin{proof}
					Without loss of generality the group $C_3$ is represented by the set of quantum channels $\{R_{0}(0), R_{0}(2\pi/3), R_{0}(4\pi/3)\}$. Consider the QSP sequence defined by QSP phase list $\Phi = \{0, -\alpha, \alpha, 0\}$ using the convention of Theorem \ref{theorem:qsp_matrix_form}, i.e., the product
						\begin{equation} \label{eq:three_angle_method}
							U_{\Phi} 
							= 
							R_{x}(\theta) R_{z}(\alpha) R_{x}(\theta) R_{z}(-\alpha) R_{x}(\theta),
						\end{equation}
					for any angle $\theta$. It is not hard to explicitly compute the top left component of this unitary operator, and specifically for the special angle $\alpha = \arccos(\cos{\theta}/[1 - \cos{\theta}])$, which is real whenever $\pi/3 \leq \theta \leq 5\pi/3$, the top left component of this unitary $\braket{0 \lvert U_\Phi \rvert 0}$ is $0$. Consequently with three queries to the oracle, and $\alpha = \arccos(-1/3)$, the transition probability $\ket{0} \mapsto \ket{0}$ is $1$ if $\theta = 0$ and $0$ if $\theta \in \{2\pi/3, 4\pi/3\}$. Consequently three additional queries are enough, possibly replacing $R_{x}(\theta)$ with $R_{x}(\theta)R_{x}(-2\pi/3)$ in (\ref{eq:three_angle_method}), to completely and perfectly determine the hidden angle. Over equal priors the expected query complexity of this technique is $5$.

					Alternatively in the language of Theorem \ref{theorem:qsp_matrix_form}, we intend that the top left element of $U_\Phi$, under the map $\cos{\theta/2} \mapsto x$, has the form
						\[
							f_1(x) = \frac{4}{3}x\,(x - 1/2)(x + 1/2),
						\]
					which is a polynomial\footnote{Note that $(4/3)(x - 1/2)(x + 1/2)$ also satisfies constraints required by QSP, and indeed this lemma can be shown using only 4 maximum (10/3 expected) oracle queries, though the resulting protocol is less geometrically obvious.} that takes modulus $1$ at $x = -1$ and $x = 1$, has definite parity, and takes value $0$ at $x = \pm 1/2$. This, along with $f_1(x)$ under the map $\theta \mapsto \theta - 2\pi/3$ produces a pair of binary measurements for which the map\footnote{$S$ is overloaded here: both channel elements and the continuous real parameter $\theta$ characterizing these elements. Note also that this map can be written as a binary tree as in Figure \ref{fig:decision_tree}.} $S \mapsto M$ is injective where $M$ is the set of binary measurements.
						\[
							\{\braket{+ \lvert U_{\Phi} \rvert +}, \braket{+ \lvert U_{\Phi}^\prime \rvert +}\}
							=
							\begin{cases}
								\{1, 0\} & \theta = 0\\
								\{0, 1\} & \theta = 2\pi/3\\
								\{0, 0\} & \theta = 4\pi/3,
							\end{cases}
						\]
					where $U^\prime_\Phi$ is the aforementioned pre-rotation replacing $R_{x}(\theta)$ with $R_{x}(\theta)R_{x}(-2\pi/3)$ or equivalently $\theta \mapsto \theta - 2\pi/3$. A visual depiction of this algorithm is given in Figure \ref{figure:three_angle_problem}, and a table relating this Lemma's construction directly to Algorithm \ref{algorithm:generic_algorithm} is given in Table \ref{fig:tabular_algorithm}.
				\end{proof}
			\end{lemma}

			\begin{table*}[htpb!]
				\begin{center}
					\begin{tabular}{*{5}{l}}
						Index  & Query map & $p_j$ & $\ket{\psi_j}$ & $\ket{\psi_j^\prime}$  \\[0.1em]
						\hline
						$j = 1$ & $\mathcal{E}_i \mapsto \mathcal{E}_i$ & $(4x^3 - x)/3$ & $\ket{+}$ & $\ket{\pm} \mapsto \{0, 1\}$\\
						$j = 2$\hspace{1em} & $\mathcal{E}_i \mapsto \mathcal{E}_iR_{x}(-2\pi/3)$\hspace{1em} & $(4x^3 - x)/3$ \hspace{1.5em} & $\ket{+}$ & $\ket{\pm} \mapsto \{0, 1\}$\\\hline\\[-0.5em]
							\multicolumn{3}{l}{
							$
							\{f_1(\mathcal{E}_i), f_2(\mathcal{E}_i)\} 
							= 
							\begin{cases}
								\{0, 1\} \mapsto R_x(0)\\
								\{1, 0\} \mapsto R_x(2\pi/3)\\
								\{1, 1\} \mapsto R_x(4\pi/3)
							\end{cases}
							$
							} 
							& 
							\multicolumn{2}{l}{(Inverse map)}
					\end{tabular}
				\end{center}
				\caption{The use of Algorithm \ref{algorithm:generic_algorithm} as a subroutine for solving the three angle problem (Lemma \ref{lemma:order_3}) in tabular form. As $p_j$ for $j \in [m]$ completely define both $f_j$ and the corresponding QSP-derived objects given in Definition \ref{def:r_qht_algorithm}, they provide a minimal explicit demonstration of the use of Algorithm \ref{algorithm:generic_algorithm}. Included are quantum states for preparation, $\ket{\psi_j}$, and measurement, $\ket{\psi_j^\prime}$, as well as the compound query map (Definition \ref{def:compound_queries}), where Algorithm \ref{algorithm:generic_algorithm} is fed compound queries. Finally, an inverse map is given to recover the hidden channel.}
				\label{fig:tabular_algorithm}
			\end{table*}

			\begin{figure}[htpb!]
				\centering
					\subfloat[Geometric interpretation.]{\includegraphics[width=0.6\columnwidth]{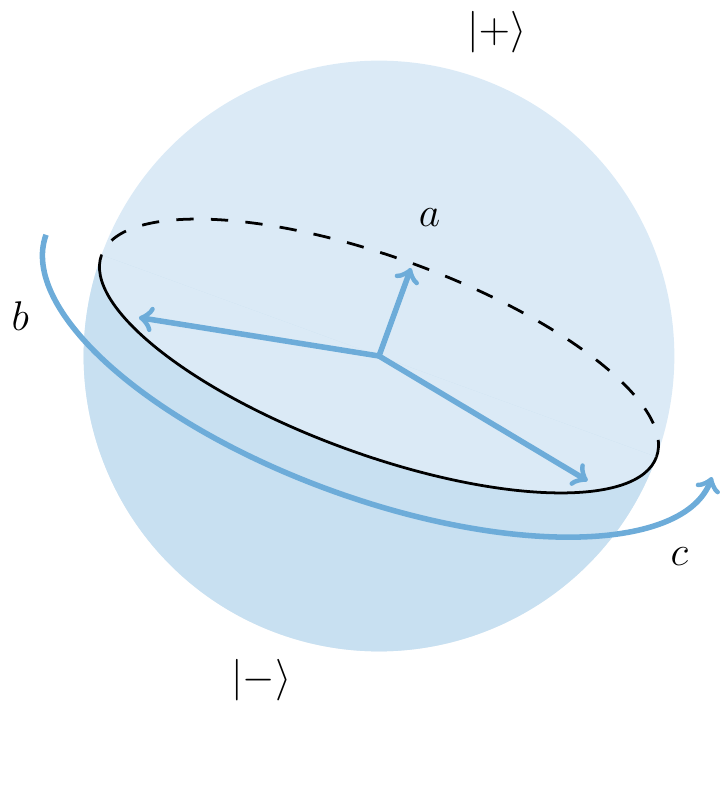}\label{subfig:geometric}}
				\hspace{1em}
					\subfloat[Algebraic interpretation.]{\includegraphics[width=0.7\columnwidth]{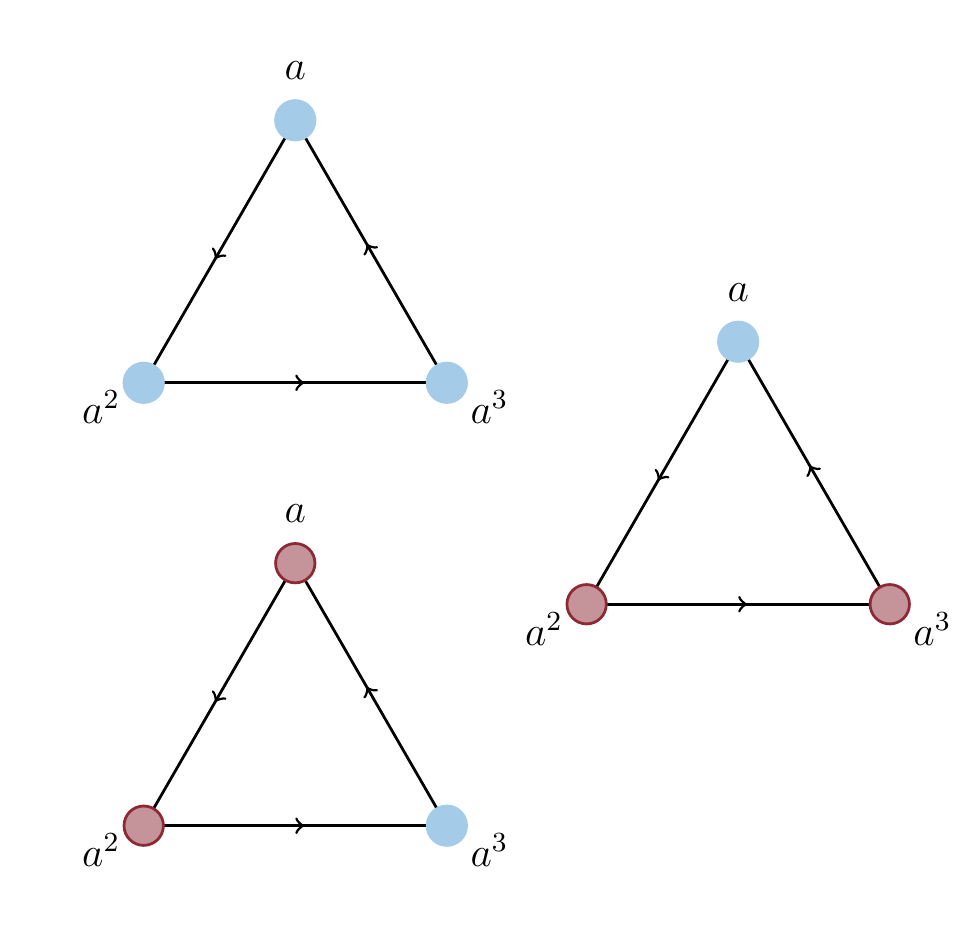}\label{subfig:algebraic}}
				\caption{Geometric (a) and algebraic (b) depictions of the proof of Lemma \ref{lemma:order_3}. Unitary representations of $C_3$ in SU(2) are, without loss of generality, equivalent to a set of rotations which cycles states $(a, b, c)$ as shown on the Bloch sphere in (a). Moving away from the Bloch sphere, any sequence of quantum channel discrimination protocols whose binary PVM output differs on subsets of quantum channels representing $C_3$ (e.g., partitions $C_3$ elements into red and blue subsets as pictured, and as proven in Lemma \ref{lemma:order_3}), also determines the queried quantum channel perfectly. The partitions indicated in (b) are generated by polynomials given in Figure \ref{figure:three_angle_problem}.}
				\label{figure:three_angle_bloch_sphere}
			\end{figure}

		The functional intuition of protocols deciding on representations of cyclic groups is depicted in Figure \ref{figure:three_angle_problem}. As discussed previously, QSP protocols take equiangular rotations about different axes in equator of the Bloch sphere (see Figure \ref{figure:three_angle_bloch_sphere}), interleave them with rotations about orthogonal axes on the Bloch sphere, and give efficient methods for forcing the corresponding matrix elements of the final, composite rotation to be desired trigonometic polynomials in the unknown rotation angle. Figure \ref{figure:three_angle_problem} demonstrates that polynomials which have modulus $0$ or $1$ at specific angles result in deterministic protocols for dividing the search space. The work remaining is to systematize sub-protocols of this form to generate efficient decision protocols on the entire query set.

		Finally we can provide a proof for perfect decision protocols on all prime order cyclic groups, and in fact this shows an even stronger result as the same method goes through for cyclic groups of any odd order. However, given the results of Lemma \ref{lemma:prime_reduction}, QSP is only a necessary tool in the prime-order case, when compound queries provide no helpful simplifications.
			
			\begin{theorem} \label{theorem:all_prime_order}
				There exists a family of deterministic algorithms $\mathcal{F} = \{\mathcal{A}_{C_p}\}$ for all primes $p$, where $\mathcal{A}_{C_p}$ perfectly decides $C_p$, with asymptotically optimal query complexity.

				\begin{proof}
					The proof follows from the existence of a family of polynomials $f_1, f_2, \cdots, f_m$ whose moduli take values in $\{0, 1\}$ on a finite set of subsets $\{S_j\}$ for $j \in [m]$ of the set of $p$ possible phases $S_0$ induced by queries to the oracle, namely
						\[
							S_0 = \left\{\cos\left(\frac{\pi n}{p}\right), \; n \in [p]\right\},
						\]
					such that that the successive subsets $S_0 \supseteq S_1 \supseteq \cdots \supseteq S_m$ have the following\footnote{Also described in Remark \ref{remark:bisection_remark} and Figure \ref{fig:decision_tree}.} properties:
						\begin{itemize}
							
							\item \textbf{Bisecting}: The order of $S_{j + 1}$ should satisfy that $\lvert S_{j + 1} \rvert \leq (1/c)\lvert S_{j} \rvert$ for some fixed constant $c = \mathcal{O}(1)$.
			                
			                \item \textbf{Density reducing}: The minimum separation between elements of $S_{j + 1}$ on which the modulus of the interpolating polynomial $f_{j + 1}$ takes distinct values should increase exponentially in $j$.
			                
			                \item \textbf{Totally deciding}: Constructing a family of interpolating polynomials $p_j$ whose moduli take values in the set $\{0, 1\}$ on $S_j$ is equivalent to computing a family of binary functions $f_j$ on $C_p$; the evaluation of these binary functions on the hidden channel corresponding to $g \in C_n$ should uniquely identify $g$. I.e., this map $g \mapsto \{0, 1\}^m$ should be injective (see Figure \ref{fig:decision_tree}).
			                
			                \item \textbf{Parity preserving}: The elements of $S_{j}$ should be of definite parity for all $j$; this parity is shared by all $p_j$.

		            	\end{itemize}

					If all of these conditions are satisfied by some judicious sequence of $S_j$ the result follows if the number of such non-trivial subsets of $S$, given by $m$, is asymptotically $\log{p}$ and the degree of the polynomial $p_j$ goes as $\mathcal{O}(p/c^j)$ in which case the entire protocol has query complexity linear in $p$.

					The existence of these interpolating polynomials is guaranteed by the results of Section \ref{section:poly_interpolation}, while their asymptotic query complexity follows directly from exponentially increasing promised gaps between elements of $S_j$. For a given group $C_p$ these subsets $S_j$ have the explicit, measurement dependent, form
						\begin{align*}
							&S_0 = S_0 \\[0.5em]
							&S_j^0 = \left\{s_k \in S_{j - 1}, \;f_{j - 1}(s_k) = 0\right\}\\[0.5em]
							&S_j^1 = \left\{s_k \in S_{j - 1}, \;f_{j - 1}(s_k) = 1\right\}
						\end{align*}
					where the new $S_j$'s upper index indicates the measurement result of the QSP sequence dividing the search space, and is subsequently dropped as this iterative division continues. The functions $f_j$ are defined as polynomials which interpolate any binary function on the set $S_{j - 1}$ which alternates maximally with definite parity on $[-1,1]$ ($f_j$ will share this parity). These functions have explicit description, e.g., when given some subset $S_j$ of size $2n + 1$, indexing by $\ell$ for increasing $s_\ell$ in $[-1,1]$.
						\[
							f_{j}(x_\ell) = 
							\begin{cases}
								\frac{1}{2}[1 + (-1)^\ell] & 1 \leq \ell \leq n\\[0.5em]
								\frac{1}{2}[1 + (-1)^{\ell-1}] & n + 1 \leq \ell \leq 2n + 1.
							\end{cases}
						\]
					This evidently preserves parity and confers the right properties on successive subsets. In plain terms this is a binary search whose constituent sub-searches grow exponentially cheaper in query complexity, and whose base case is handled by Lemma \ref{lemma:order_3}.
				\end{proof}
			\end{theorem}
		
		Finally, by the previous results we can make a statement for all cyclic groups, and proceed to richer subgroups of SU(2).
			
			\begin{corollary} \label{corollary:all_cyclic_groups}
				For all $n \in \mathbb{N}$, there exists a deterministic algorithm $\mathcal{A}_{C_n}$ which perfectly decides $C_n$, with asymptotically optimal query complexity. This follows directly from Lemma \ref{lemma:prime_reduction} and Theorem \ref{theorem:all_prime_order}.
			\end{corollary}

	\subsection{Dihedral groups} \label{subsection:dihedral_group}

\begin{figure}[htpb!]
				\begin{center}
					\includegraphics[width=1.0\columnwidth]{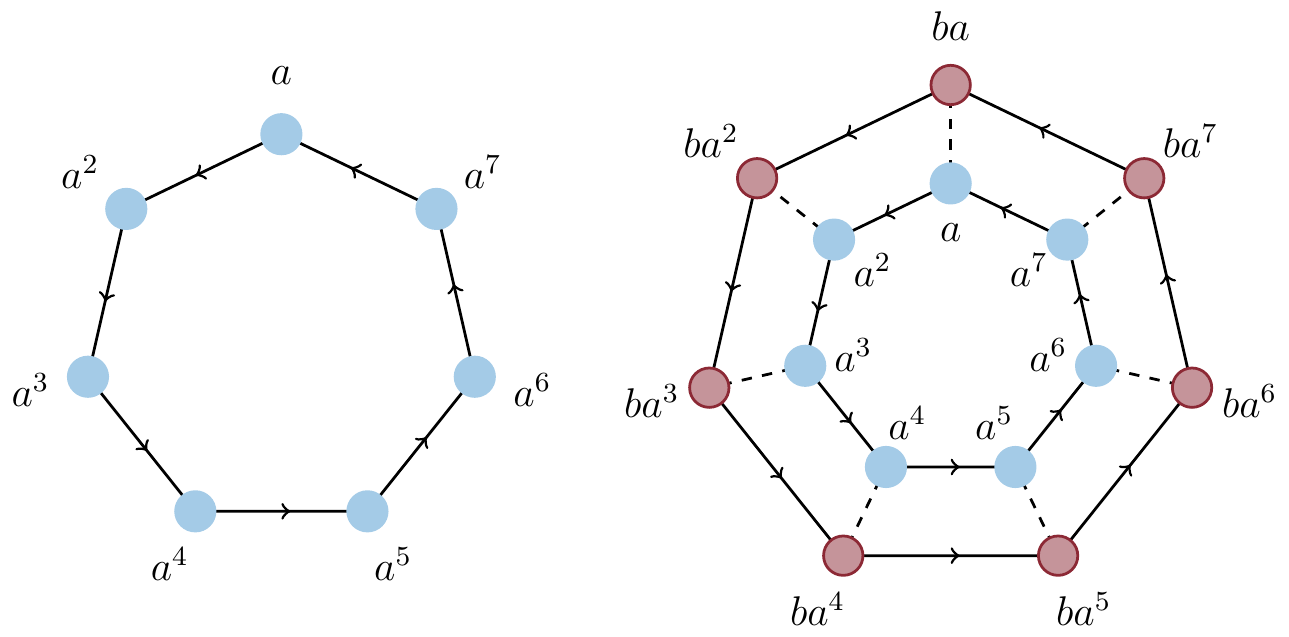}
				\end{center}
	  			\caption{Two presentations of Cayley graphs for the cyclic group of order 7 and the dihedral group of order 14. The observation that the cyclic group admits no irreducible representations in SU(2) allows the perfect determination, in one additional query, of coset membership for the maximal cyclic subgroup of $D_{2n}$, partitioning the red and blue sets.}
	  			\label{figure:cayley_groups}
			\end{figure}

		We consider the dihedral groups of order $2n$; it is not too difficult to see that each bit-string label for an element $g \in D_{2n}$ requires exactly one more bit to uniquely describe the element, corresponding to membership of $g$ in one of two cosets of the normal cyclic subgroup $C_n \triangleleft D_n$. We show that this bit can be recovered in one additional measurement, and that our protocol is thus optimal assuming the optimality of the protocol which decides $C_n$.
			\begin{theorem} \label{theorem:dihedral_decision}
				Assuming existence of an algorithm $\mathcal{A}_{C_n}$ that perfectly decides $C_n$ there exists an algorithm $\mathcal{A}_{D_{2n}}$ that calls $\mathcal{A}_{C_n}$ as a sub-routine and perfectly decides $D_{2n}$, the dihedral group of order $2n$, with one additional oracle query. A depiction of the overarching idea of this algorithm is given in Figure \ref{figure:cayley_groups}.

				\begin{proof}
					Without loss of generality $\mathcal{A}_{D_{2n}}$ has oracle access to a channel in a representation of $D_{2n}$ whose cyclic subgroup $C_n$ in SU(2) has representation:
						\begin{equation}
							\{R_{z}(m\cdot 2\pi/n)\}, \; m \in [n].
						\end{equation}
					The SU(2) embedding of $D_{2n}$ that contains our embedding of $C_n$ as a subgroup is generated by a generator of this $C_n$, $\sigma$, and another group element, $\tau$, which without loss of generality has representation $R_x(\pi)$. The standard presentation of $D_{2n}$ is
						\begin{equation}
							D_{2n} \equiv \{\sigma, \tau \;|\; \sigma^n = \tau^2 = \tau\sigma\tau\sigma = e\},
						\end{equation}
					The lemma follows if there exists a simple protocol to, given query access to an unknown element $g \in D_{2n}$, determine membership of the queried element $g$ among the two cosets of $C_n < D_{2n}$.

					Assume $U(g)$ is the unitary operation corresponding to the group element $g$ embedded in SU(2) as stated. Then
						\begin{align}
							H U(\braket{\sigma}) H\ket{0} &= \ket{0}\\[0.5em]
							H U(\tau) U(\braket{\sigma}) H\ket{0} &= \ket{1},
						\end{align}
					where $H$ is the Hadamard gate and $U(\braket{\sigma})$ represents some unitary operation within the subgroup $\braket{\sigma}$ generated by $\sigma$. Intuitively, $H$ rotates $\ket{0}$ to another state insensitive to the action of the cyclic index $2$ subgroup of $D_{2n}$. This follows from the lack of irreducible representations of $C_n$ in SU(2).

					If $\ket{0}$ is measured then $\mathcal{A}_{C_n}$ can be applied as normal to future queries, respecting the embedding of the subgroup $\braket{\sigma}$. Otherwise any query made to the oracle $U(g)$ is prefaced by $R_x(\pi)$, reducing to a decision on $C_n$. Only one additional query is needed by $\mathcal{A}_{D_{2n}}$ to decide $D_{2n}$ (a group with twice the size).
				\end{proof}
			\end{theorem}
			
			\begin{table*}[htpb!]
				\begin{center}
					\begin{tabular}{*{5}{l}}
						Index  & Query map & $p_j$ & $\ket{\psi_j}$ & $\ket{\psi_j^\prime}$\\[0.1em]
						\hline
						$j = 1$ & $\mathcal{E}_i \mapsto R_{\xi}(-\pi/2) \mathcal{E}_i R_{\xi}(\pi/2)$\hspace{1em} & $x$ & $\ket{+}$ & $\ket{\pm} \mapsto \{0, 1\}$\\
						\multicolumn{1}{c}{\vdots} &\multicolumn{1}{c}{\vdots} &\multicolumn{1}{l}{\vdots} &\multicolumn{1}{l}{\vdots} &\multicolumn{1}{c}{\vdots} 
						\\\hline\\[-0.5em]
							\multicolumn{3}{l}{
							$
							\{f_1(\mathcal{E}_i), f_2(\mathcal{E}_i)\} 
							= 
							\begin{cases}
								\{0, \cdots\} \mapsto e * C_n^?\\
								\{1, \cdots\} \mapsto \tau * C_n^?
							\end{cases}
							$
							} 
							& 
							\multicolumn{2}{l}{(Inverse map)}
					\end{tabular}
				\end{center}
				\caption{The use of Algorithm \ref{algorithm:generic_algorithm} as a subroutine for deciding on dihedral groups $D_{2n}$ (Theorem \ref{theorem:dihedral_decision}) in tabular form. The table proceeds until reduction to $C_n$ is achieved (i.e., after the first query); this query rotates to the basis in which $\sigma$ (the generator of $C_n \triangleleft D_{2n}$) acts trivially on $\{\ket{\pm}\}$. Once coset membership in the maximal cyclic subgroup of the queried element is known, it can be inverted and applied to form compound queries that reduce the query set to $C_n$, given in Corollary \ref{corollary:all_cyclic_groups}. Here $C_n^?$ is an unknown power of $\sigma$.}
				\label{fig:tabular_algorithm_dihedral}
			\end{table*}

	\subsection{Platonic groups}

		Finally we address protocols for the finite subgroups of SU(2) that do not fit into countably infinite families, and exhibit a richer non-abelian structure than the dihedral group. These are often referred to as the platonic groups due to their appearance in the study of symmetry groups of platonic solids. Before discussing protocols for deciding $A_4$, $S_4$ and $A_5$ we define two basic group theoretic concepts that will aid in their construction.

		\begin{definition} \label{def:cycle_decomposition}
			(Cycle decomposition). Let $S$ be a finite set, e.g., the integers $\{1, 2, \cdots, n\}$, and $\sigma$ a permutation $S \rightarrow S$. The \emph{cycle decomposition} of $\sigma$ expresses $\sigma$ as a product of disjoint cycles. For instance, if $S$ has size $4$ and the action of $\sigma$ swaps pairs $1, 2$ and $3, 4$, then the cycle decomposition of $\sigma$ is denoted $(1, 2)(3, 4)$, where the order of tuples is not uniquely defined.
		\end{definition}

		\begin{definition}
			(Cycle type). Let $S$ be a finite set, for instance the integers $\{1, 2, \cdots, n\}$, and $\sigma$ a permutation $S \rightarrow S$. The \emph{cycle type}\footnote{The cycle type is sometimes defined as a tuple of the lengths of each cycle in the cycle decomposition, rather than the number of cycles of each given length.} of $\sigma$ is a tuple indicating the number of cycles of each given length in the \emph{cycle decomposition} of $\sigma$. E.g., for the example given in Definition \ref{def:cycle_decomposition}, the cycle decomposition $(1, 2)(3, 4)$ has cycle type $(0, 2, 0, 0)$, indicating two length-two cycles.

			Note that the set of all possible cycle types is in bijection with unordered partitions of the integers in $\{1, 2, \cdots, n\}$: i.e., for cycle type tuple $c$, the sum of $c_j \cdot j$ for $j \in [n]$ is simply $n$.
		\end{definition}

		\begin{theorem} \label{theorem:alternating_algorithm}
			There exists a deterministic algorithm $\mathcal{A}_{A_4}$ that perfectly decides $A_4$ with asymptotically optimal query complexity. This algorithm is additionally given in Table \ref{fig:tabular_algorithm_alternating}.
			\begin{proof}
				The elements of $A_4$ can be classified according to their cycle type as permutations on four elements. For $A_4$ these types are $(1, 0, 1, 0)$, $(0, 2, 0, 0)$ and $(4, 0, 0, 0)$ (the last being the identity permutation). 

				Cubes of any element in $A_4$ have cycle type $(0, 2, 0, 0)$ or $(4, 0, 0, 0)$ only, meaning that if the queried element $g$ is already in one of three representations for the $D_4$ normal subgroup of $A_4$ then running the $D_4$ algorithm on cubes of physical query elements gives the correct answer, and otherwise acts as if the queried element were the identity. This element can be determined in at most three compound queries deterministically, measuring in three mutually unbiased bases on the Bloch sphere, corresponding to the eigenstates of each of the generators of the chosen $D_4$ subgroup.

				Given that $A_4 \equiv D_4 \rtimes C_3$, all elements $g$ can be written in the form $kh$ where $h$ is drawn from a chosen normal $D_4$ subgroup of the representation of $A_4$ and $k$ is from a realized $C_3$ subgroup. By pre-applying powers of a generator of one of these $C_3$ subgroups, the $D_4$ algorithm on cubes of queries will consistently compute the binary function of membership of the queried element $g$ in a particular coset of the normal $D_4 \triangleleft A_4$. Assuming equal priors, such an algorithm is expected to\footnote{The explicit calculation is $(1/4)\cdot6 + (1/4)\cdot12 + (1/4)\cdot18 + (1/4)\cdot22 = 29/2$ for the $3 + 3 + 3$ non-trivial elements of the cosets of the normal $D_4$ followed by $3$ trivial elements.} terminate in 14.5 queries.
			\end{proof}
		\end{theorem}
		
		\begin{definition}
		    We give a name to a particular subroutine presented in Theorem \ref{theorem:alternating_algorithm}, whose use is indicated in Table \ref{fig:tabular_algorithm_alternating}.
		    
		    The function \texttt{correctCoset} takes as input the evaluation of the three binary measurements given in Theorem \ref{theorem:alternating_algorithm} to determine which element of the $D_4$ normal subgroup of $A_4$ enters into the chosen semi-direct product $D_4 \rtimes C_3$ and returns the representation of the inverse of this element.
		\end{definition}
		
		\begin{table*}
    		\begin{center}
    			\begin{tabular}{*{5}{l}}
    				Index & Query map & $p_j$ & $\ket{\psi_j}$ & $\ket{\psi_j^\prime}$\\[0.1em]
    				\hline
    				$j = 1$ & $\mathcal{E}_i \mapsto (\mathcal{E}_i)^3$ & $x$ & $\ket{+}$ & $\ket{\pm} \mapsto \{0, 1\}$\\
    				$j = 2$ & $\mathcal{E}_i \mapsto R_{x}(\pi/2)(\mathcal{E}_i)^3R_{x}(-\pi/2)$ & $x$ & $\ket{+}$ & $\ket{\pm} \mapsto \{0, 1\}$\\
    				$j = 3$ & $\mathcal{E}_i \mapsto R_{x^\prime}(\pi/2)(\mathcal{E}_i)^3R_{x^\prime}(-\pi/2)$ & $x$ & $\ket{+}$ & $\ket{\pm} \mapsto \{0, 1\}$\\
    				$j = 4$ & $\mathcal{E}_i \mapsto \mathcal{E}_i\texttt{correctCoset}(f_{< 4}(\mathcal{E}_i))$ & $(4x^3 - x)/3$\hspace{1em} & $\ket{+}$ & $\ket{\pm} \mapsto \{0, 1\}$\\
    				$j = 5$ & $\mathcal{E}_i \mapsto R_{x^{\prime\prime}}(2\pi/3)\mathcal{E}_i\texttt{correctCoset}(f_{< 4}(\mathcal{E}_i))$\hspace{1em} & $(4x^3 - x)/3$ & $\ket{+}$ & $\ket{\pm} \mapsto \{0, 1\}$\\\hline\\[-0.5em]
    					\multicolumn{3}{l}{
    					$
    					\{f_1(\mathcal{E}_i), f_2(\mathcal{E}_i), f_3(\mathcal{E}_i),\cdots\} 
    					= 
    					\begin{cases}
    						\{0, 0, 0, \cdots\} \mapsto \mask{D_4^{ab}}{D_4^{e}} * C_3^{?}\\
    						\{1, 1, 0, \cdots\} \mapsto \mask{D_4^{ab}}{D_4^{a}} * C_3^{?}\\
    						\{1, 0, 1, \cdots\} \mapsto \mask{D_4^{ab}}{D_4^{b}} * C_3^{?}\\
    						\{0, 1, 1, \cdots\} \mapsto D_4^{ab} * C_3^{?}
    					\end{cases}
    					$
    					} 
    					& 
    					\multicolumn{2}{c}{(Inverse map)}
    			\end{tabular}
    		\end{center}
    		\caption{The use of Algorithm \ref{algorithm:generic_algorithm} as a subroutine for deciding on $A_4$ as in Theorem \ref{theorem:alternating_algorithm} in tabular form. As $p_j$ for $j \in [m]$ completely define both $f_j$ and the corresponding QSP-derived objects given in Definition \ref{def:r_qht_algorithm}, they provide a minimal explicit demonstration of the use of Algorithm \ref{algorithm:generic_algorithm}. The $p_j$ given here have also had their QSP angles explicitly given in Lemma \ref{lemma:order_3}. Here $D^{g}_{4}*C^{?}_3$ indicates a group element in the semi-direct product defining $A_4$ which is the product of $g$, an element of the chosen $D_4$ normal subgroup in terms of generators $\{a, b\}$ and an unknown element of the chosen $C_3$ subgroup. Axes $x, x^\prime$ are chosen such that these rotations generate the chosen $D_4$ subgroup, and $x^{\prime\prime}$ the axis of rotation for the chosen $C_3$.}
    		\label{fig:tabular_algorithm_alternating}
    	\end{table*}

		\begin{theorem}
			There exists a deterministic algorithm $\mathcal{A}_{S_4}$ which perfectly decides $S_4$, with asymptotically optimal query complexity.
			\begin{proof}
				Squares of elements in $S_4$ necessarily fall in the alternating group $A_4$, though this mapping is not always invertible. It is invertible, however, when the queried element $g$ in $S_4$ has the cycle type $(1, 0, 1, 0)$. For any element in $S_4$ there exists an element $h$ of cycle type $(2, 1, 0, 0)$ for which the product $gh$ is of cycle type $(1, 0, 1, 0)$. Consequently there exists an algorithm that, for every element $h$ of cycle type $(2, 1, 0, 0)$, of which there are six, pre-applies $h$ to queries (and repeats this process to generate squares of this query element, $ghgh = (gh)^2$) and runs the $A_4$ algorithm on this compound query, which recovers perfectly in finitely many queries the hidden element $g$ when the image $(gh)^2$ has cycle type $(1, 0, 1, 0)$. Namely there exists a subroutine which determines coset membership for cosets of the normal $A_4 \triangleleft S_4$, and proceeds by reduction to decision on $A_4$. This protocol is expected to terminate in 34 queries.\footnote{Again this number is arrived at by explicitly writing a table of elements of $S_4$ and running them through the protocol as given until it terminates.}
			\end{proof}
		\end{theorem}

		We note that the two protocols given above make no reference to the mechanisms of QSP, but are instead completely algebraic in form, exploiting the simple canonical subgroup towers of $A_4$ and $S_4$ to reduce decisions on representations of these groups to those on their normal subgroups. It is the small size of these non-abelian groups in particular which, unfortunately, bring the following remark. Resolving this problem is left open as stated in Section \ref{section:conclusions}.

		\begin{remark}
			The alternating group on five elements has, unfortunately, no simple reduction to an algorithm of the previous, smaller groups, in part because $A_5$ is the smallest simple non-abelian group, and thus permits no non-trivial decompositions in terms of a canonical tower of subgroups.
		\end{remark}
		
		Before concluding this section we give an overview (Remark \ref{remark:recursion_remark}) of the major technique which has permitted the extension of algorithms solving R-QHT (i.e., Algorithm \ref{algorithm:generic_algorithm}) to those solving G-QHT.
		
        \begin{remark} \label{remark:recursion_remark}
            Extending the recursive bisection depicted in Figure \ref{fig:decision_tree}, which in turn demonstrates the methods of Algorithm \ref{algorithm:generic_algorithm}, to representations of non-cyclic groups follows, in each instance described in this section, from the following sketched protocol.
                
    		For each finite group presented in Section \ref{section:explicit_construction}, we must provide (1) a small quantum circuit to produce compound queries (Definition \ref{def:compound_queries}) satisfying the input assumptions of Algorithm \ref{algorithm:generic_algorithm}, (2) apply Algorithm \ref{algorithm:generic_algorithm} and keep track of its minimal required query complexity, and finally (3) verify the satisfaction of conditions under which the compound query map is invertible (e.g., as in Remark \ref{remark:bisection_remark}), these conditions remaining unchanged despite the introduction of compound queries.
    		
    		Whether this protocol is possible to perform for general groups is an open question, and indeed the methods of this section relied on the fact that the finite groups investigated were non-simple and often semi-direct products of abelian groups.
        \end{remark}

\section{Extending QHT protocols to larger groups and noisy settings} \label{section:generalizations}

    It is natural to consider generalizations to the setting in which the results of Section \ref{section:explicit_construction} were derived. This section concerns itself with two: (1) the inclusion of noise and (2) extension to larger finite groups.
    
    \subsection{Noisy channels and noisy quantum gates}

		The algorithms of Section \ref{section:explicit_construction} relied on the fact that compound queries to the oracle could, under the assumption of access to unitary channels, make perfect use of the algebraic relations which were a priori known among the query set. These relations led to \emph{effective query access} to simpler query sets for whom the optimal hypothesis testing algorithm was known. Naturally, however, realistic quantum computers and quantum channels exhibit noise, and one might be concerned about two different sources of error as summarized below.

			\begin{enumerate}

				\item The queried elements $U_g$ may not perfectly satisfy the conditions imposed on a faithful representation of $G$, but may instead only \emph{approximately} satisfy them, i.e.,
					\begin{align*}
						U_{g}U_{h} \approx_{\epsilon} U_{gh} \quad \forall g, h \in G,
					\end{align*}
				where the approximate equality is with respect to some reasonable norm, here the diamond norm. Alternatively one can consider that the channels themselves are only near-unitary, i.e., that $U_g^\prime \approx_\epsilon U_g$ for all $g \in G$ where the norm is again reasonable. Such a channel might be given by its operator-sum representation
					\[
						U_g \equiv \int_{h} f_g(h)U_h\,d\mu(h),
					\]
				where $f_g(h)$ is some probability density function defined suitably on elements of SU(2) which is peaked about $g$ to induce an operator whose diamond norm with $U_g$ is suitably small. Here $\mu$ is some suitable measure over SU(2).

				\item The unitary operators applied by the querent may, in general, also not be perfect. This is the statement that the rotations normally applied in a QSP sequence as per the statement of Algorithm \ref{algorithm:generic_algorithm} may again only satisfy $U_j^\prime \approx_\epsilon U_j$ for all indices $j$ in the QSP sequence. We denote by $U_j^\prime$ the applied unitary and by $U_j$ the intended unitary.

			\end{enumerate}

		We consider the first instance, namely the physically realistic scenario that the ideal query set $S$ is not the sampled query set, but instead that physical queries may be slightly perturbed from ideal queries. I.e., the physical queries $\{\mathcal{E}_i^\prime\}$ are such that the diamond distance $\lVert \mathcal{E}_i - \mathcal{E}^\prime_i \rVert_{\diamond} \leq \epsilon$ for some small $\epsilon > 0$. In this case, which encompasses all small perturbations, methods analogous to the `peeling lemma' in \cite{pirandola}, permit us to bound our new error in discrimination.
			\begin{lemma}
				Fixing an initial state $\rho_j$ the trace distance $\lVert \rho - \rho^\prime \rVert$ between the serial quantum channel discrimination protocol defined by the interspersed unitaries $\{U_{i,j}\}$ acting on $\rho_j$ where the queried channel set is $\{\mathcal{E}_i\}$ versus $\{\mathcal{E}_i^\prime\}$ is bounded above by $n_j \lVert \mathcal{E}_m - \mathcal{E}_m^\prime \rVert_{\diamond} \leq n_j \epsilon$.

				\begin{proof}
					In the case that the QSP sequences used are length 2, we show the result, and show that the method generalizes to length $n_j$ sequences. The distance $\lVert \rho - \rho^\prime \rVert$ can be re\"expressed and bounded above according to
						\begin{align*}
							&\phantom{{}={}} \lVert 
							U_2 \circ \mathcal{E}_m \circ U_1 \circ  \mathcal{E}_m(\rho_j) 
							- 
							U_2 \circ \mathcal{E}_m^\prime \circ U_1 \circ  \mathcal{E}_m^\prime(\rho_j) 
							\rVert\\
							&\leq  \lVert 
							\mathcal{E}_m \circ U_1 \circ  \mathcal{E}_m(\rho_j) 
							- 
							\mathcal{E}_m^\prime \circ U_1 \circ  \mathcal{E}_m^\prime(\rho_j) 
							\rVert\\
							&\leq  \lVert 
							\mathcal{E}_m \circ U_1 \circ  \mathcal{E}_m(\rho_j) 
							- 
							\mathcal{E}_m \circ U_1 \circ  \mathcal{E}_m^\prime(\rho_j) 
							\rVert
							\\&\hspace{2em}+
							\lVert 
							\mathcal{E}_m^\prime \circ U_1 \circ  \mathcal{E}_m(\rho_j) 
							- 
							\mathcal{E}_m^\prime \circ U_1 \circ  \mathcal{E}_m^\prime(\rho_j) 
							\rVert\\
							&\leq 
							\lVert \mathcal{E}_m(\rho_j) - \mathcal{E}_m^\prime(\rho_j) \rVert 
							\\&\hspace{2em}+
							\lVert \mathcal{E}_m[U_1 \circ \mathcal{E}^\prime(\rho_j)] 
							- \mathcal{E}_m^\prime[U_1 \circ \mathcal{E}_m^\prime(\rho_j)] \rVert\\
							&\leq 2\lVert \mathcal{E}_m - \mathcal{E}_m^\prime \rVert_{\diamond},
						\end{align*}
					where the inequalities, in order from top to bottom, follow from (1) the monotonicity of the trace distance, (2) the triangle inequality, (3) monotonicity with respect to the CPTP map $\mathcal{E}_m^\prime\circ U_1$, and (4) that the diamond distance dominates the trace distance on any particular initial state. This result can be iterated for arbitrarily many channel applications where the coefficient on the diamond distance goes as $n_j$ where $n_j$ is the discrimination algorithm's $j$-th subpart's query complexity.
				\end{proof}
			\end{lemma}

		For the second instance, where the querent's own unitary operations are only close to the ideal operations, an analogous argument to that used in \cite{kitaev} permits us to bound error to a multiple of the per-gate error $\epsilon$ (usually computed in terms of a trace distance between the intended and applied channel) where this multiple is proportional to the QSP sequence's length. Consequently under reasonable assumptions of noise in both the queried channel and the locally applied unitary operators, the methods presented in the previous section do no worse than expected, accruing error linearly in sequence length for reasonable norms.

	\subsection{Extensions to larger groups} \label{subsection:larger_groups}

		The methods of Section \ref{section:explicit_construction} use \emph{compound queries} (e.g., positive integer powers of queries), defined in Problem \ref{extended_p_qht_problem}, to access representations of subgroups of $G$. It is thus of interest to determine when one is to expect that (1) subsets of $m$-th powers of group elements generate proper subgroups, and (2) what information can be extracted under the assumption of the ability to decide on said subgroups. We state a series of related lemmas regarding these questions, assuming a basic understanding of group theory.

		The following two lemmas in particular discuss sufficient conditions under which a known normal subgroup of $G$ permits query access to compound queries that reside in said normal subgroup. These lemmas capture the underlying mechanism of the protocols given previously for deciding $D_{2n}$ and $A_4$.

		\begin{lemma} \label{lemma:m_power_groups}
			If a finite group $G$ admits a normal subgroup $N$ of index $m$ then the subset of $m$-th powers of $G$, equivalently $S^m = \{g_1^m, g_2^m, \cdots, g_n^m \}$ for all $n$ elements of $G$ generates a proper subgroup $G^\prime \leq N < G$.

			\begin{proof}
				Proof follows from recognizing that elements of the form $g_i^m$ are in the kernel of the group homomorphism $G \rightarrow G/N$ and thus $\langle S^m \rangle$ is a (possibly non-proper) subgroup of the normal subgroup $N$ of $G$, equivalently $\langle S^m \rangle \leq N < G$.
			\end{proof}
		\end{lemma}

		\begin{lemma} \label{lemma:m_power_groups_structure}
			If a finite group $G$ admits a normal subgroup $N$ of index $m$ then the subset of $m$-th powers of $G$, i.e., the group generated by $S^m$ as in Lemma \ref{lemma:m_power_groups}, is contained within the intersection of all index $m$ normal subgroups of $G$. Proof again follows by the isomorphism theorems.
		\end{lemma}

		Furthermore we give a lemma that describes the underlying behavior of the protocol given previously for deciding on $S_4$ (Theorem \ref{theorem:alternating_algorithm}). It is the statement of this lemma, as well as the two preceding it, that precludes a solution for deciding on $A_5$, which admits no non-trivial normal subgroups.

		\begin{lemma} \label{lemma:m_power_invertible}
			If the $m$-power map $g \mapsto g^m$ applied to elements of $G$ generates a proper subgroup $G^\prime < G$, and there exists a group element $h \in G$ such that for some subset $S$ of $G$ the map $s \mapsto (sh)^m$ is invertible for all $s \in S$, and there exists a quantum protocol for deciding $G^\prime$, then there exists a quantum protocol for deciding the query set $G^\prime \cup S$.

			\begin{proof}
				Constructing compound queries of $m$-th powers of physical queries allows access (at $m$ times the query complexity) to a representation of $G^\prime$. The statement of the lemma with respect to the set $S$ says merely that pre-application of $h \in G$ before each query $s$ is, under the map given, invertible, and thus unique identification of $s$ is also possible with knowledge of $h$.
			\end{proof}
		\end{lemma}

		The statements given in the lemmas above do not depend on particularly complicated notions in group theory; instead, we have simply asked which simple operations can be performed in our limited resource model to faithfully simplify the query set. In most instances, these simplifications correspond to the existence of normal subgroups (equivalently kernels of group homomorphisms). For statements beyond those given here, especially those concerning the conditions under which the assumptions of Lemma \ref{lemma:m_power_invertible} hold beyond $S_4$, we define a selection of open problems in Section \ref{section:conclusions}.

		The procedure outlined in Lemma \ref{lemma:m_power_invertible} is also not the most general one; indeed, compound queries can be built from general products of known unitary operations (some of which may coincide with the query set) and possibly multiple copies the queried channel. Conditions under which such a map is invertible relate intimately to the study of characters in representation theory, and provide exciting avenues for improved protocols for larger finite groups, e.g., $G < SU(n)$. Moreover, when considering larger Hilbert spaces, in analogy to the algorithms deciding on the dihedral groups $D_{2n}$, the family of finite groups which permit no irreducible representation in said larger Hilbert space grows richer, and correspondingly decisions on groups which are semi-direct products grow easier. Thus, while extension to larger Hilbert spaces may not resolve the discussion of efficient decision algorithms on all larger groups in the serial adaptive query model, it may reasonably result in interesting \emph{quantitative} statements on the entanglement or auxiliary system size necessary to achieve efficient (query-complexity-wise) discrimination dependent on the nature of the represented group.

\section{Discussion and conclusions} \label{section:conclusions}

	In this work we have provided a constructive approach for achieving efficient quantum multiple hypothesis testing for query sets whose algebra faithfully represents a finite subgroup of SU(2). The nature of this construction centers on the use of Algorithm \ref{algorithm:generic_algorithm}, a quantum algorithm for solving the simpler R-QHT problem (Problem \ref{r_qht_problem}), as a subroutine along with methods for exploiting known algebraic structure of the query set to enable reductions to R-QHT. This reduction is summarized in Remark \ref{remark:recursion_remark}.

	Concretely, when the represented group $G$ is either abelian or both non-abelian and non-simple the protocols we construct achieve optimal query complexity without the use of auxiliary systems or entanglement; this statement is equivalent to a statement about the minimal degree of constrained interpolating polynomials, and resolves an open question in \cite{duan_feng_ying}, as well as generalizes an old result in \cite{davies} to quantum channels. Moreover, the bridge that Algorithm \ref{algorithm:generic_algorithm} and its derivate algorithms demonstrate between quantum information and functional approximation theory indicates a rich variety of novel instantiations of the basic ideas of QSP \cite{low-chuang, gilyen}.

	In addition to achieving efficient quantum channel discrimination for a family of channel sets in a serial adaptive query model, we show that our protocols can be aborted early while still accomplishing useful tasks; this follows simply from the nature of the binary search discussed in Remark \ref{remark:bisection_remark}. For instance, the reductions provided throughout Section \ref{section:explicit_construction} are directly realizable as coset membership testing procedures in general, or period finding for the case of cyclic groups.
	
	In the following remarks and problem definitions, we provide one more direct application of the methods discussed in this work to a problem in quantum communication.
	
	    \begin{remark} \label{remark:reference_remark}
	        As mentioned in \cite{chiribella_05, bit_commitment_07}, efficient protocols for the estimation of unitary processes have use in the transmission of reference frames as well as various proofs of insecurity for device-independent protocols for quantum bit-commitment.
	        
	        We give one example for how this work can be applied to a discretized version (e.g., group frames \cite{group_frames, frame_introduction}, which share close relation with SIC-POVMs) of reference frame-sharing (Problem \ref{problem:discrete_reference_frame} and Lemma \ref{lemma:discrete_reference_frame}).
	    \end{remark}
	    
	    \begin{problem} \label{problem:discrete_reference_frame}
	        Consider two separated parties, Alice and Bob; each is able to (1) perform single-qubit unitaries and (2) transmit qubits noiselessly to the other party. Alice and Bob agree on a shared $z$-axis but are rotated with respect to each other by some unknown angle $\theta$ about this axis. Moreover, the possible $\theta$ lie within a discrete set $\Theta$ of size $n$, known to both parties.

	        Alice and Bob can come to agreement on the unknown angle $\theta$ with certainty in a finite length interactive protocol; this protocol is denoted \emph{dual QSP} due to its similarities with \emph{standard QSP} \cite{low-16, low-19, low-chuang, gilyen}, and is said to solve the \emph{dual QSP problem}.
	    \end{problem}
	    
	    \begin{lemma} \label{lemma:discrete_reference_frame}
	        There exists a finite length interactive interactive protocol by which two parties playing the game defined in Problem \ref{problem:discrete_reference_frame} can win with certainty and with asymptotically optimal round number (under the restriction of sending single qubits).
	        
	        \begin{proof}
                Proof proceeds by direct construction. Beginning with some initial state $\ket{\psi_0}$, Alice applies to it a rotation about her local $x$ axis, namely $\exp{(i\phi_0\sigma_x)}$, and sends this qubit to Bob. Bob applies a rotation about \emph{his} local $x$-axis by another specified angle $\phi_1$, or equivalently according to Alice (if she knew the angle $\theta$) Bob appears to apply $\exp{(i\phi_1[\cos{\theta}\sigma_x + \sin{\theta}\sigma_y])} = U_B \exp{(i\phi_1\sigma_x)} U_B^{-1}$ where $U_B = \exp{(-i[\theta/2]\sigma_z)}$.
                
                In other words, Alice and Bob can, according to some previously agreed upon prescription of real angles $\Phi = \{\phi_0, \phi_1, \cdots, \phi_m\}$, \emph{collaboratively compute} the unitary operator\footnote{Here assuming that $m$ is even, i.e., that the protocol ends with Alice receiving the qubit.}
    				\begin{eqnarray}
    					U_\Phi = e^{i\phi_m\sigma_x} \cdots &&e^{-i[\theta/2]\sigma_z} e^{i\phi_3\sigma_x} e^{i[\theta/2]\sigma_z} e^{i\phi_2\sigma_x}\nonumber\\ &&e^{-i[\theta/2]\sigma_z} e^{i\phi_1\sigma_x} e^{i[\theta/2]\sigma_z} e^{i\phi_0\sigma_x}.
    				\end{eqnarray}
    			Moreover, following the final application of $\exp{(i\phi_m\sigma_x)}$ and measurement against $\ket{\psi_{1}}$, Alice can sample from the Bernoulli distribution defined by the transition probability $p = \lvert\braket{\psi_1 \lvert U_\Phi \rvert \psi_0}\rvert^2$.
    			
    			The construction above is almost a vanilla QSP sequence. It is not so difficult to see that if Alice and Bob additionally apply the rotation $\exp\{\pm i[\pi/2]\sigma_x\}$ respectively, locally, after their $\phi_j$ rotation for $j \in \{1,2, \cdots, m\}$, the collaborative sequence instead becomes
            		\begin{eqnarray}
            			U_{\Phi^\prime} = e^{i\phi_m\sigma_x} \cdots &&e^{i[\theta/2]\sigma_z} e^{i\phi_3\sigma_x} e^{i[\theta/2]\sigma_z}
            			e^{i\phi_2\sigma_x}\nonumber\\  &&e^{i[\theta/2]\sigma_z} e^{i\phi_1\sigma_x} e^{i[\theta/2]\sigma_z} e^{i\phi_0\sigma_x},
            		\end{eqnarray}
            	which is of the form of a standard QSP sequence. Consequently we see concrete connection between \emph{dual QSP} and standard QSP: i.e., a redefinition of QSP phase angles.
            	
            	Given a standard QSP strategy, defined by an angular sequence $\Phi$, there exists an angular sequence $\Phi^\prime$ following the prescription given above such that the \emph{dual QSP} sequence defined by $\Phi^\prime$ acts identically given access to parties of relative angular displacement $\theta$ as the sequence defined by $\Phi$ acts given query access to an equiangular rotation $\exp{(-i[\theta/2]\sigma_z)}$ in the setting of standard QSP.
            	
            	Consequently a protocol solving Problem \ref{problem:discrete_reference_frame} follows directly from a protocol solving Problem \ref{r_qht_problem} under the prescription (following Algorithm \ref{algorithm:generic_algorithm}) defined by $\Phi_{j, k}^{\prime} = \Phi_{j, k} + \pi$ for $k \in \{1, \cdots, n_j\}$ and $\Phi_{j, 0}^{\prime} = \Phi_{j, 0}$.
	        \end{proof}
	    \end{lemma}
	    
	    \begin{remark}
	        We can analyze the performance of the protocol given in Lemma \ref{lemma:discrete_reference_frame} in two ways: (1) in comparison na\"ive repetition of binary hypothesis testing and (2) in comparison to phase estimation, the continuous analogue of the problem statement.
                \begin{itemize}
                    \item The results of \cite{duan_feng_ying} assert that the query complexity for distinguishing two distinct unitary operators $U, V$ scales as $\mathcal{O}(1/\Theta[U^\dag V])$ where $\Theta[W]$ is the length of the smallest arc containing all the eigenvalues of $W$ on the unit circle in the complex plane.
	                
	                When phrased as a decision on a representation\footnote{This merely connects $n$ in a reasonable, i.e., reciprocal, functional map to a factor defining the difficulty of discrimination, in which the stated quadratic improvement is always possible.} of $C_n$, eliminating one possible quantum channel at a time gives a query complexity that scales as $\mathcal{O}(n^2)$ (as $\mathcal{O}(n)$ such discrimination procedures are required, each costing $\mathcal{O}(n)$ queries). As shown in the constructions leading to Corollary \ref{corollary:all_cyclic_groups}, however, decisions on $C_n$ and consequently also discrete reference-frame sharing, have query complexity scaling as $\mathcal{O}(n)$ (up to logarithmic factors) courtesy of the implicit binary search in Algorithm \ref{algorithm:generic_algorithm}.
	                
                    \item A feature of Lemma \ref{lemma:discrete_reference_frame} is that it yields a deterministic quantum algorithm. If one only wishes to determine the relative rotation with high confidence, one can use phase estimation and achieve the same $\mathcal{O}(n)$ query complexity scaling \cite{nielsen_chuang} using $\mathcal{O}(\log{n} + \log{(1/\epsilon)})$ qubits for confidence $\epsilon$. This also matches the performance of the estimation procedure in \cite{chiribella_05}. Thus while estimative methods perform similarly in the cyclic group case to G-QHT-derived methods, the methodology of Lemma \ref{lemma:discrete_reference_frame} is tailored to the statement of discrete reference-frame sharing, can be done serially, and can be extended to richer finite groups.
                \end{itemize}
        \end{remark}

    The methods of Lemma \ref{lemma:discrete_reference_frame} suggest a useful technique; namely, whenever a suitable sensing problem can be (1) discretized and (2) made coherent, the ability to, by a simple quantum process, induce a phase on, e.g., a single qubit, allows all of the mechanisms built in earlier sections to be directly applied with concomitant statements about query complexity or round complexity\footnote{In the methods given, query complexity and round complexity are precisely the same (under the map from dual QSP to standard QSP): transmission of the shared qubit is necessary to enact a unitary operation dependent on the relative rotation.} optimality.

	Beyond direct applicability to discrete versions of problems defined in prior work (e.g., reference frame sharing), several fundamental open problems remain whose solution might lie in methods related to those discussed in this work; we outline a few of them below.

		\begin{itemize}

			\item \textbf{Decisions on the subgroup tower}: In analogy to the protocol given for deciding the dihedral group in Subsection \ref{subsection:dihedral_group}, are there families of larger groups $G^\prime$ whose lack of irreducible representation in the natural Hilbert space of multiple qubits ($[\mathbb{C}^{2}]^{\otimes n}$) or qudits ($\mathbb{C}^d$) permits groups $G$ whose canonical subgroup tower includes $G^\prime$ to be decided by reduction to decisions on $G^\prime$? What are sufficient conditions under which protocols deciding $G$ can, even inefficiently, be reduced to protocols for deciding normal subgroups of $G$? Small examples of this phenomenon are given in the lemmas of Subsection \ref{subsection:larger_groups}.

			\item \textbf{Optimal G-QHT with bounded entanglement}: Given the procedure in the above part, does there exist a quantifiable trade-off between the serial and parallel query model query complexities required for deciding groups $G$ given access to Hilbert spaces in which no representation of $G$ is irreducible? If entanglement is required for optimal QHT algorithms on large or highly non-abelian query sets, are there methods to quantify the required minimum entanglement?

			\item \textbf{Quantum property testing}: Do there exist partial discrimination protocols, e.g., beyond those provided for deciding coset membership, which decide other interesting properties of the group represented by the query set while not totally deciding on the group?

			\item \textbf{Estimating compact group elements}: Can the performance of quantum channel estimation protocols for compact groups $G$, e.g., as in \cite{chiribella_05}, be suitably recovered by employing a method similar to those of this work to systematically divide the search space up to within a specified error? Under what assumptions about the compact group is this decision-to-estimation conversion in the serial adaptive query model still efficient?

		\end{itemize}

	To summarize, major avenues for extending this work lie in (1) natural generalizations to higher dimensional Hilbert spaces and (2) characterizations of richer finite groups which find natural representations in higher dimensional Hilbert spaces. Improvements in methods to address these questions have implications in quantum algorithms for problems in discrete algebra, and this subfield in turn has potential application, following translation of G-QHT-like problems to novel contexts (e.g., as in Lemma \ref{lemma:discrete_reference_frame}), to useful quantum algorithms for cryptography, communication, and sensing.
	
\section{Acknowledgments} \label{acknowledgements}

	This work was supported in part by the NSF Center for Ultracold Atoms (CUA), the NSF EPiQC Expedition in Computing, and NTT Research.

\nocite{*}

\bibliography{main}

\begin{thebibliography}{44}%
\makeatletter
\providecommand \@ifxundefined [1]{%
 \@ifx{#1\undefined}
}%
\providecommand \@ifnum [1]{%
 \ifnum #1\expandafter \@firstoftwo
 \else \expandafter \@secondoftwo
 \fi
}%
\providecommand \@ifx [1]{%
 \ifx #1\expandafter \@firstoftwo
 \else \expandafter \@secondoftwo
 \fi
}%
\providecommand \natexlab [1]{#1}%
\providecommand \enquote  [1]{``#1''}%
\providecommand \bibnamefont  [1]{#1}%
\providecommand \bibfnamefont [1]{#1}%
\providecommand \citenamefont [1]{#1}%
\providecommand \href@noop [0]{\@secondoftwo}%
\providecommand \href [0]{\begingroup \@sanitize@url \@href}%
\providecommand \@href[1]{\@@startlink{#1}\@@href}%
\providecommand \@@href[1]{\endgroup#1\@@endlink}%
\providecommand \@sanitize@url [0]{\catcode `\\12\catcode `\$12\catcode
  `\&12\catcode `\#12\catcode `\^12\catcode `\_12\catcode `\%12\relax}%
\providecommand \@@startlink[1]{}%
\providecommand \@@endlink[0]{}%
\providecommand \url  [0]{\begingroup\@sanitize@url \@url }%
\providecommand \@url [1]{\endgroup\@href {#1}{\urlprefix }}%
\providecommand \urlprefix  [0]{URL }%
\providecommand \Eprint [0]{\href }%
\providecommand \doibase [0]{https://doi.org/}%
\providecommand \selectlanguage [0]{\@gobble}%
\providecommand \bibinfo  [0]{\@secondoftwo}%
\providecommand \bibfield  [0]{\@secondoftwo}%
\providecommand \translation [1]{[#1]}%
\providecommand \BibitemOpen [0]{}%
\providecommand \bibitemStop [0]{}%
\providecommand \bibitemNoStop [0]{.\EOS\space}%
\providecommand \EOS [0]{\spacefactor3000\relax}%
\providecommand \BibitemShut  [1]{\csname bibitem#1\endcsname}%
\let\auto@bib@innerbib\@empty
\bibitem [{\citenamefont {Helstrom}(1976)}]{helstrom}%
  \BibitemOpen
  \bibfield  {author} {\bibinfo {author} {\bibfnamefont {C.~W.}\ \bibnamefont
  {Helstrom}},\ }\href@noop {} {\emph {\bibinfo {title} {Quantum Detection and
  Estimation Theory}}},\ Mathematics in Science and Engineering: A Series of
  Monographs and Textbooks\ (\bibinfo  {publisher} {Academic Press, Inc.},\
  \bibinfo {year} {1976})\BibitemShut {NoStop}%
\bibitem [{\citenamefont {Pirandola}\ \emph {et~al.}(2019)\citenamefont
  {Pirandola}, \citenamefont {Laurenza}, \citenamefont {Lupo},\ and\
  \citenamefont {Pereira}}]{pirandola}%
  \BibitemOpen
  \bibfield  {author} {\bibinfo {author} {\bibfnamefont {S.}~\bibnamefont
  {Pirandola}}, \bibinfo {author} {\bibfnamefont {R.}~\bibnamefont {Laurenza}},
  \bibinfo {author} {\bibfnamefont {C.}~\bibnamefont {Lupo}},\ and\ \bibinfo
  {author} {\bibfnamefont {J.~L.}\ \bibnamefont {Pereira}},\ }\bibfield
  {title} {\bibinfo {title} {Fundamental limits to quantum channel
  discrimination},\ }\href
  {https://doi.org/https://doi.org/10.1038/s41534-019-0162-y} {\bibfield
  {journal} {\bibinfo  {journal} {npj Quantum Inf.}\ }\textbf {\bibinfo
  {volume} {5}},\ \bibinfo {pages} {50} (\bibinfo {year} {2019})}\BibitemShut
  {NoStop}%
\bibitem [{\citenamefont {Ac{\'{i}}n}(2001)}]{acin}%
  \BibitemOpen
  \bibfield  {author} {\bibinfo {author} {\bibfnamefont {A.}~\bibnamefont
  {Ac{\'{i}}n}},\ }\bibfield  {title} {\bibinfo {title} {Statistical
  distinguishability between unitary operations},\ }\href
  {https://doi.org/10.1103/PhysRevLett.87.177901} {\bibfield  {journal}
  {\bibinfo  {journal} {Phys. Rev. Lett.}\ }\textbf {\bibinfo {volume} {87}},\
  \bibinfo {pages} {177901} (\bibinfo {year} {2001})}\BibitemShut {NoStop}%
\bibitem [{\citenamefont {Duan}\ \emph {et~al.}(2009)\citenamefont {Duan},
  \citenamefont {Feng},\ and\ \citenamefont {Ying}}]{duan}%
  \BibitemOpen
  \bibfield  {author} {\bibinfo {author} {\bibfnamefont {R.}~\bibnamefont
  {Duan}}, \bibinfo {author} {\bibfnamefont {Y.}~\bibnamefont {Feng}},\ and\
  \bibinfo {author} {\bibfnamefont {M.}~\bibnamefont {Ying}},\ }\bibfield
  {title} {\bibinfo {title} {Perfect distinguishability of quantum
  operations},\ }\href {https://doi.org/10.1103/PhysRevLett.103.210501}
  {\bibfield  {journal} {\bibinfo  {journal} {Phys. Rev. Lett.}\ }\textbf
  {\bibinfo {volume} {103}},\ \bibinfo {pages} {210501} (\bibinfo {year}
  {2009})}\BibitemShut {NoStop}%
\bibitem [{\citenamefont {Duan}\ \emph {et~al.}(2007)\citenamefont {Duan},
  \citenamefont {Feng},\ and\ \citenamefont {Ying}}]{duan_feng_ying}%
  \BibitemOpen
  \bibfield  {author} {\bibinfo {author} {\bibfnamefont {R.}~\bibnamefont
  {Duan}}, \bibinfo {author} {\bibfnamefont {Y.}~\bibnamefont {Feng}},\ and\
  \bibinfo {author} {\bibfnamefont {M.}~\bibnamefont {Ying}},\ }\bibfield
  {title} {\bibinfo {title} {Entanglement is not necessary for perfect
  discrimination between unitary operations},\ }\href
  {https://doi.org/10.1103/PhysRevLett.98.100503} {\bibfield  {journal}
  {\bibinfo  {journal} {Phys. Rev. Lett.}\ }\textbf {\bibinfo {volume} {98}},\
  \bibinfo {pages} {100503} (\bibinfo {year} {2007})}\BibitemShut {NoStop}%
\bibitem [{\citenamefont {Zhuang}\ and\ \citenamefont
  {Pirandola}(2020)}]{zhuang_pirandola}%
  \BibitemOpen
  \bibfield  {author} {\bibinfo {author} {\bibfnamefont {Q.}~\bibnamefont
  {Zhuang}}\ and\ \bibinfo {author} {\bibfnamefont {S.}~\bibnamefont
  {Pirandola}},\ }\bibfield  {title} {\bibinfo {title} {Ultimate limits for
  multiple quantum channel discrimination},\ }\href
  {https://doi.org/10.1103/PhysRevLett.125.080505} {\bibfield  {journal}
  {\bibinfo  {journal} {Phys. Rev. Lett.}\ }\textbf {\bibinfo {volume} {125}},\
  \bibinfo {pages} {080505} (\bibinfo {year} {2020})}\BibitemShut {NoStop}%
\bibitem [{\citenamefont {Hashimoto}\ \emph {et~al.}(2010)\citenamefont
  {Hashimoto}, \citenamefont {Hayashi}, \citenamefont {Hayashi},\ and\
  \citenamefont {Horibe}}]{hashimoto_10}%
  \BibitemOpen
  \bibfield  {author} {\bibinfo {author} {\bibfnamefont {T.}~\bibnamefont
  {Hashimoto}}, \bibinfo {author} {\bibfnamefont {A.}~\bibnamefont {Hayashi}},
  \bibinfo {author} {\bibfnamefont {M.}~\bibnamefont {Hayashi}},\ and\ \bibinfo
  {author} {\bibfnamefont {M.}~\bibnamefont {Horibe}},\ }\bibfield  {title}
  {\bibinfo {title} {Unitary-process discrimination with error margin},\ }\href
  {https://doi.org/10.1103/PhysRevA.81.062327} {\bibfield  {journal} {\bibinfo
  {journal} {Phys. Rev. A}\ }\textbf {\bibinfo {volume} {81}},\ \bibinfo
  {pages} {062327} (\bibinfo {year} {2010})}\BibitemShut {NoStop}%
\bibitem [{\citenamefont {Takagi}\ \emph {et~al.}(2019)\citenamefont {Takagi},
  \citenamefont {Regula}, \citenamefont {Bu}, \citenamefont {Liu},\ and\
  \citenamefont {Adesso}}]{takagi}%
  \BibitemOpen
  \bibfield  {author} {\bibinfo {author} {\bibfnamefont {R.}~\bibnamefont
  {Takagi}}, \bibinfo {author} {\bibfnamefont {B.}~\bibnamefont {Regula}},
  \bibinfo {author} {\bibfnamefont {K.}~\bibnamefont {Bu}}, \bibinfo {author}
  {\bibfnamefont {Z.}~\bibnamefont {Liu}},\ and\ \bibinfo {author}
  {\bibfnamefont {G.}~\bibnamefont {Adesso}},\ }\bibfield  {title} {\bibinfo
  {title} {Operational advantage of quantum resources in subchannel
  discrimination},\ }\href {https://doi.org/10.1103/PhysRevLett.122.140402}
  {\bibfield  {journal} {\bibinfo  {journal} {Phys. Rev. Lett.}\ }\textbf
  {\bibinfo {volume} {122}},\ \bibinfo {pages} {140402} (\bibinfo {year}
  {2019})}\BibitemShut {NoStop}%
\bibitem [{\citenamefont {Takagi}\ and\ \citenamefont
  {Regula}(2019)}]{takagi_general}%
  \BibitemOpen
  \bibfield  {author} {\bibinfo {author} {\bibfnamefont {R.}~\bibnamefont
  {Takagi}}\ and\ \bibinfo {author} {\bibfnamefont {B.}~\bibnamefont
  {Regula}},\ }\bibfield  {title} {\bibinfo {title} {General resource theories
  in quantum mechanics and beyond: Operational characterization via
  discrimination tasks},\ }\href {https://doi.org/10.1103/PhysRevX.9.031053}
  {\bibfield  {journal} {\bibinfo  {journal} {Phys. Rev. X}\ }\textbf {\bibinfo
  {volume} {9}},\ \bibinfo {pages} {031053} (\bibinfo {year}
  {2019})}\BibitemShut {NoStop}%
\bibitem [{\citenamefont {Harrow}\ \emph {et~al.}(2010)\citenamefont {Harrow},
  \citenamefont {Hassidim}, \citenamefont {Leung},\ and\ \citenamefont
  {Watrous}}]{harrow-10}%
  \BibitemOpen
  \bibfield  {author} {\bibinfo {author} {\bibfnamefont {A.~W.}\ \bibnamefont
  {Harrow}}, \bibinfo {author} {\bibfnamefont {A.}~\bibnamefont {Hassidim}},
  \bibinfo {author} {\bibfnamefont {D.~W.}\ \bibnamefont {Leung}},\ and\
  \bibinfo {author} {\bibfnamefont {J.}~\bibnamefont {Watrous}},\ }\bibfield
  {title} {\bibinfo {title} {Adaptive versus nonadaptive strategies for quantum
  channel discrimination},\ }\href {https://doi.org/10.1103/PhysRevA.81.032339}
  {\bibfield  {journal} {\bibinfo  {journal} {Phys. Rev. A}\ }\textbf {\bibinfo
  {volume} {81}},\ \bibinfo {pages} {032339} (\bibinfo {year}
  {2010})}\BibitemShut {NoStop}%
\bibitem [{\citenamefont {Sacchi}(2005)}]{sacchi_05}%
  \BibitemOpen
  \bibfield  {author} {\bibinfo {author} {\bibfnamefont {M.~F.}\ \bibnamefont
  {Sacchi}},\ }\bibfield  {title} {\bibinfo {title} {Optimal discrimination of
  quantum operations},\ }\href {https://doi.org/10.1103/PhysRevA.71.062340}
  {\bibfield  {journal} {\bibinfo  {journal} {Phys. Rev. A}\ }\textbf {\bibinfo
  {volume} {71}},\ \bibinfo {pages} {062340} (\bibinfo {year}
  {2005})}\BibitemShut {NoStop}%
\bibitem [{\citenamefont {Davies}(1978)}]{davies}%
  \BibitemOpen
  \bibfield  {author} {\bibinfo {author} {\bibfnamefont {E.}~\bibnamefont
  {Davies}},\ }\bibfield  {title} {\bibinfo {title} {Information and quantum
  measurement},\ }\href {https://doi.org/10.1109/TIT.1978.1055941} {\bibfield
  {journal} {\bibinfo  {journal} {IEEE Trans. Inf. Theory}\ }\textbf {\bibinfo
  {volume} {24}},\ \bibinfo {pages} {596} (\bibinfo {year} {1978})}\BibitemShut
  {NoStop}%
\bibitem [{\citenamefont {Kuperberg}(2005)}]{kuperberg}%
  \BibitemOpen
  \bibfield  {author} {\bibinfo {author} {\bibfnamefont {G.}~\bibnamefont
  {Kuperberg}},\ }\bibfield  {title} {\bibinfo {title} {A subexponential-time
  quantum algorithm for the dihedral hidden subgroup problem},\ }\href
  {https://doi.org/https://doi.org/10.1137/S0097539703436345} {\bibfield
  {journal} {\bibinfo  {journal} {SIAM J. Comput.}\ }\textbf {\bibinfo {volume}
  {35}},\ \bibinfo {pages} {170} (\bibinfo {year} {2005})}\BibitemShut
  {NoStop}%
\bibitem [{\citenamefont {Regev}(2004)}]{regev}%
  \BibitemOpen
  \bibfield  {author} {\bibinfo {author} {\bibfnamefont {O.}~\bibnamefont
  {Regev}},\ }\bibfield  {title} {\bibinfo {title} {Quantum computation and
  lattice problems},\ }\href
  {https://doi.org/https://doi.org/10.1137/S0097539703440678} {\bibfield
  {journal} {\bibinfo  {journal} {SIAM J. Comput.}\ }\textbf {\bibinfo {volume}
  {33}},\ \bibinfo {pages} {738–} (\bibinfo {year} {2004})}\BibitemShut
  {NoStop}%
\bibitem [{\citenamefont {Childs}\ and\ \citenamefont {van
  Dam}(2010)}]{childs_10}%
  \BibitemOpen
  \bibfield  {author} {\bibinfo {author} {\bibfnamefont {A.~M.}\ \bibnamefont
  {Childs}}\ and\ \bibinfo {author} {\bibfnamefont {W.}~\bibnamefont {van
  Dam}},\ }\bibfield  {title} {\bibinfo {title} {Quantum algorithms for
  algebraic problems},\ }\href {https://doi.org/10.1103/revmodphys.82.1}
  {\bibfield  {journal} {\bibinfo  {journal} {Rev. Mod. Phys.}\ }\textbf
  {\bibinfo {volume} {82}},\ \bibinfo {pages} {1} (\bibinfo {year}
  {2010})}\BibitemShut {NoStop}%
\bibitem [{\citenamefont {Chiribella}\ \emph {et~al.}(2005)\citenamefont
  {Chiribella}, \citenamefont {D'Ariano},\ and\ \citenamefont
  {Sacchi}}]{chiribella_05}%
  \BibitemOpen
  \bibfield  {author} {\bibinfo {author} {\bibfnamefont {G.}~\bibnamefont
  {Chiribella}}, \bibinfo {author} {\bibfnamefont {G.~M.}\ \bibnamefont
  {D'Ariano}},\ and\ \bibinfo {author} {\bibfnamefont {M.~F.}\ \bibnamefont
  {Sacchi}},\ }\bibfield  {title} {\bibinfo {title} {Optimal estimation of
  group transformations using entanglement},\ }\href
  {https://doi.org/10.1103/PhysRevA.72.042338} {\bibfield  {journal} {\bibinfo
  {journal} {Phys. Rev. A}\ }\textbf {\bibinfo {volume} {72}},\ \bibinfo
  {pages} {042338} (\bibinfo {year} {2005})}\BibitemShut {NoStop}%
\bibitem [{\citenamefont {Waldron}(2013)}]{group_frames}%
  \BibitemOpen
  \bibfield  {author} {\bibinfo {author} {\bibfnamefont {S.}~\bibnamefont
  {Waldron}},\ }\bibfield  {title} {\bibinfo {title} {Group frames},\ }in\
  \href@noop {} {\emph {\bibinfo {booktitle} {Finite Frames, Theory and
  Applications}}},\ \bibinfo {editor} {edited by\ \bibinfo {editor}
  {\bibfnamefont {P.~G.}\ \bibnamefont {Casazza}}\ and\ \bibinfo {editor}
  {\bibfnamefont {G.}~\bibnamefont {Kutyniok}}}\ (\bibinfo  {publisher}
  {Birkh{\"{a}}user},\ \bibinfo {address} {New York},\ \bibinfo {year} {2013})\
  Chap.~\bibinfo {chapter} {5}, pp.\ \bibinfo {pages} {171--192}\BibitemShut
  {NoStop}%
\bibitem [{\citenamefont {Kovačević}\ and\ \citenamefont
  {Chebira}(2008)}]{frame_introduction}%
  \BibitemOpen
  \bibfield  {author} {\bibinfo {author} {\bibfnamefont {J.}~\bibnamefont
  {Kovačević}}\ and\ \bibinfo {author} {\bibfnamefont {A.}~\bibnamefont
  {Chebira}},\ }\bibfield  {title} {\bibinfo {title} {An introduction to
  frames},\ }\href {https://doi.org/10.1561/2000000006} {\bibfield  {journal}
  {\bibinfo  {journal} {Found. Trends Signal Process.}\ }\textbf {\bibinfo
  {volume} {2}},\ \bibinfo {pages} {1} (\bibinfo {year} {2008})}\BibitemShut
  {NoStop}%
\bibitem [{\citenamefont {Chiribella}\ \emph {et~al.}(2008)\citenamefont
  {Chiribella}, \citenamefont {D'Ariano},\ and\ \citenamefont
  {Perinotti}}]{memory_channels_08}%
  \BibitemOpen
  \bibfield  {author} {\bibinfo {author} {\bibfnamefont {G.}~\bibnamefont
  {Chiribella}}, \bibinfo {author} {\bibfnamefont {G.~M.}\ \bibnamefont
  {D'Ariano}},\ and\ \bibinfo {author} {\bibfnamefont {P.}~\bibnamefont
  {Perinotti}},\ }\bibfield  {title} {\bibinfo {title} {Memory effects in
  quantum channel discrimination},\ }\href
  {https://doi.org/10.1103/PhysRevLett.101.180501} {\bibfield  {journal}
  {\bibinfo  {journal} {Phys. Rev. Lett.}\ }\textbf {\bibinfo {volume} {101}},\
  \bibinfo {pages} {180501} (\bibinfo {year} {2008})}\BibitemShut {NoStop}%
\bibitem [{\citenamefont {D'Ariano}\ \emph {et~al.}(2007)\citenamefont
  {D'Ariano}, \citenamefont {Kretschmann}, \citenamefont {Schlingemann},\ and\
  \citenamefont {Werner}}]{bit_commitment_07}%
  \BibitemOpen
  \bibfield  {author} {\bibinfo {author} {\bibfnamefont {G.~M.}\ \bibnamefont
  {D'Ariano}}, \bibinfo {author} {\bibfnamefont {D.}~\bibnamefont
  {Kretschmann}}, \bibinfo {author} {\bibfnamefont {D.}~\bibnamefont
  {Schlingemann}},\ and\ \bibinfo {author} {\bibfnamefont {R.~F.}\ \bibnamefont
  {Werner}},\ }\bibfield  {title} {\bibinfo {title} {Reexamination of quantum
  bit commitment: The possible and the impossible},\ }\href
  {https://doi.org/10.1103/PhysRevA.76.032328} {\bibfield  {journal} {\bibinfo
  {journal} {Phys. Rev. A}\ }\textbf {\bibinfo {volume} {76}},\ \bibinfo
  {pages} {032328} (\bibinfo {year} {2007})}\BibitemShut {NoStop}%
\bibitem [{\citenamefont {Low}\ \emph {et~al.}(2016)\citenamefont {Low},
  \citenamefont {Yoder},\ and\ \citenamefont {Chuang}}]{low-16}%
  \BibitemOpen
  \bibfield  {author} {\bibinfo {author} {\bibfnamefont {G.~H.}\ \bibnamefont
  {Low}}, \bibinfo {author} {\bibfnamefont {T.~J.}\ \bibnamefont {Yoder}},\
  and\ \bibinfo {author} {\bibfnamefont {I.~L.}\ \bibnamefont {Chuang}},\
  }\bibfield  {title} {\bibinfo {title} {Methodology of resonant equiangular
  composite quantum gates},\ }\href {https://doi.org/10.1103/PhysRevX.6.041067}
  {\bibfield  {journal} {\bibinfo  {journal} {Phys. Rev. X}\ }\textbf {\bibinfo
  {volume} {6}},\ \bibinfo {pages} {041067} (\bibinfo {year}
  {2016})}\BibitemShut {NoStop}%
\bibitem [{\citenamefont {Low}\ and\ \citenamefont
  {Chuang}(2017)}]{low-chuang}%
  \BibitemOpen
  \bibfield  {author} {\bibinfo {author} {\bibfnamefont {G.~H.}\ \bibnamefont
  {Low}}\ and\ \bibinfo {author} {\bibfnamefont {I.~L.}\ \bibnamefont
  {Chuang}},\ }\bibfield  {title} {\bibinfo {title} {Optimal hamiltonian
  simulation by quantum signal processing},\ }\href
  {https://doi.org/10.1103/PhysRevLett.118.010501} {\bibfield  {journal}
  {\bibinfo  {journal} {Phys. Rev. Lett.}\ }\textbf {\bibinfo {volume} {118}},\
  \bibinfo {pages} {010501} (\bibinfo {year} {2017})}\BibitemShut {NoStop}%
\bibitem [{\citenamefont {Low}\ and\ \citenamefont {Chuang}(2019)}]{low-19}%
  \BibitemOpen
  \bibfield  {author} {\bibinfo {author} {\bibfnamefont {G.~H.}\ \bibnamefont
  {Low}}\ and\ \bibinfo {author} {\bibfnamefont {I.~L.}\ \bibnamefont
  {Chuang}},\ }\bibfield  {title} {\bibinfo {title} {Hamiltonian simulation by
  qubitization},\ }\href {https://doi.org/10.22331/q-2019-07-12-163} {\bibfield
   {journal} {\bibinfo  {journal} {Quantum}\ }\textbf {\bibinfo {volume} {3}},\
  \bibinfo {pages} {163} (\bibinfo {year} {2019})}\BibitemShut {NoStop}%
\bibitem [{\citenamefont {Gily\'{e}n}\ \emph {et~al.}(2019)\citenamefont
  {Gily\'{e}n}, \citenamefont {Su}, \citenamefont {Low},\ and\ \citenamefont
  {Wiebe}}]{gilyen}%
  \BibitemOpen
  \bibfield  {author} {\bibinfo {author} {\bibfnamefont {A.}~\bibnamefont
  {Gily\'{e}n}}, \bibinfo {author} {\bibfnamefont {Y.}~\bibnamefont {Su}},
  \bibinfo {author} {\bibfnamefont {G.~H.}\ \bibnamefont {Low}},\ and\ \bibinfo
  {author} {\bibfnamefont {N.}~\bibnamefont {Wiebe}},\ }\bibfield  {title}
  {\bibinfo {title} {Quantum singular value transformation and beyond:
  exponential improvements for quantum matrix arithmetics},\ }in\ \href
  {https://doi.org/10.1145/3313276.3316366} {\emph {\bibinfo {booktitle}
  {Proceedings of the 51st Annual ACM SIGACT Symposium on Theory of
  Computing}}},\ \bibinfo {series and number} {STOC 2019}\ (\bibinfo
  {publisher} {Association for Computing Machinery},\ \bibinfo {year} {2019})\
  p.\ \bibinfo {pages} {193–204}\BibitemShut {NoStop}%
\bibitem [{\citenamefont {Harrow}\ \emph {et~al.}(2009)\citenamefont {Harrow},
  \citenamefont {Hassidim},\ and\ \citenamefont {Lloyd}}]{harrow-09}%
  \BibitemOpen
  \bibfield  {author} {\bibinfo {author} {\bibfnamefont {A.~W.}\ \bibnamefont
  {Harrow}}, \bibinfo {author} {\bibfnamefont {A.}~\bibnamefont {Hassidim}},\
  and\ \bibinfo {author} {\bibfnamefont {S.}~\bibnamefont {Lloyd}},\ }\bibfield
   {title} {\bibinfo {title} {Quantum algorithm for linear systems of
  equations},\ }\href {https://doi.org/10.1103/PhysRevLett.103.150502}
  {\bibfield  {journal} {\bibinfo  {journal} {Phys. Rev. Lett.}\ }\textbf
  {\bibinfo {volume} {103}},\ \bibinfo {pages} {150502} (\bibinfo {year}
  {2009})}\BibitemShut {NoStop}%
\bibitem [{\citenamefont {Dong}\ \emph {et~al.}(2020)\citenamefont {Dong},
  \citenamefont {Meng}, \citenamefont {Whaley},\ and\ \citenamefont
  {Lin}}]{dmwl_20}%
  \BibitemOpen
  \bibfield  {author} {\bibinfo {author} {\bibfnamefont {Y.}~\bibnamefont
  {Dong}}, \bibinfo {author} {\bibfnamefont {X.}~\bibnamefont {Meng}}, \bibinfo
  {author} {\bibfnamefont {B.}~\bibnamefont {Whaley}},\ and\ \bibinfo {author}
  {\bibfnamefont {L.}~\bibnamefont {Lin}},\ }\bibfield  {title} {\bibinfo
  {title} {Efficient phase factor evaluation in quantum signal processing}}
  (\bibinfo {year} {2020}),\ \bibinfo {note} {arXiv Preprint.
  \url{https://arxiv.org/abs/2002.11649}}\BibitemShut {NoStop}%
\bibitem [{\citenamefont {Lin}\ and\ \citenamefont {Tong}(2019)}]{lin-tong}%
  \BibitemOpen
  \bibfield  {author} {\bibinfo {author} {\bibfnamefont {L.}~\bibnamefont
  {Lin}}\ and\ \bibinfo {author} {\bibfnamefont {Y.}~\bibnamefont {Tong}},\
  }\bibfield  {title} {\bibinfo {title} {Solving quantum linear system problem
  with near-optimal complexity}} (\bibinfo {year} {2019}),\ \bibinfo {note}
  {arXiv Preprint. \url{https://arxiv.org/abs/1910.14596}}\BibitemShut
  {NoStop}%
\bibitem [{\citenamefont {Haah}(2019)}]{haah_19}%
  \BibitemOpen
  \bibfield  {author} {\bibinfo {author} {\bibfnamefont {J.}~\bibnamefont
  {Haah}},\ }\bibfield  {title} {\bibinfo {title} {Product decomposition of
  periodic functions in quantum signal processing},\ }\href
  {https://doi.org/10.22331/q-2019-10-07-190} {\bibfield  {journal} {\bibinfo
  {journal} {Quantum}\ }\textbf {\bibinfo {volume} {3}},\ \bibinfo {pages}
  {190} (\bibinfo {year} {2019})}\BibitemShut {NoStop}%
\bibitem [{\citenamefont {Erdos}(1943)}]{erdos}%
  \BibitemOpen
  \bibfield  {author} {\bibinfo {author} {\bibfnamefont {P.}~\bibnamefont
  {Erdos}},\ }\bibfield  {title} {\bibinfo {title} {On some convergence
  properties of the interpolation polynomials},\ }\href@noop {} {\bibfield
  {journal} {\bibinfo  {journal} {Ann. Math}\ }\textbf {\bibinfo {volume}
  {44}},\ \bibinfo {pages} {330} (\bibinfo {year} {1943})}\BibitemShut
  {NoStop}%
\bibitem [{\citenamefont {Wolibner}(1951)}]{wolibner}%
  \BibitemOpen
  \bibfield  {author} {\bibinfo {author} {\bibfnamefont {W.}~\bibnamefont
  {Wolibner}},\ }\bibfield  {title} {\bibinfo {title} {Sur un polynome
  d’interpolation},\ }\href@noop {} {\bibfield  {journal} {\bibinfo
  {journal} {Colloquium Mathematicae}\ }\textbf {\bibinfo {volume} {2}},\
  \bibinfo {pages} {136} (\bibinfo {year} {1951})}\BibitemShut {NoStop}%
\bibitem [{\citenamefont {Mhaskar}\ \emph {et~al.}(2001)\citenamefont
  {Mhaskar}, \citenamefont {Narcowich}, \citenamefont {Sivakumar},\ and\
  \citenamefont {Ward}}]{mhaskar-sain}%
  \BibitemOpen
  \bibfield  {author} {\bibinfo {author} {\bibfnamefont {H.~N.}\ \bibnamefont
  {Mhaskar}}, \bibinfo {author} {\bibfnamefont {F.~J.}\ \bibnamefont
  {Narcowich}}, \bibinfo {author} {\bibfnamefont {N.}~\bibnamefont
  {Sivakumar}},\ and\ \bibinfo {author} {\bibfnamefont {J.~D.}\ \bibnamefont
  {Ward}},\ }\bibfield  {title} {\bibinfo {title} {Approximation with
  interpolatory constraints},\ }\href@noop {} {\bibfield  {journal} {\bibinfo
  {journal} {Proc. Am. Math. Soc.}\ }\textbf {\bibinfo {volume} {130}},\
  \bibinfo {pages} {1355} (\bibinfo {year} {2001})}\BibitemShut {NoStop}%
\bibitem [{\citenamefont {McLaughlin}\ and\ \citenamefont
  {Zaretzki}(1971)}]{mclaughlin-sain}%
  \BibitemOpen
  \bibfield  {author} {\bibinfo {author} {\bibfnamefont {H.~W.}\ \bibnamefont
  {McLaughlin}}\ and\ \bibinfo {author} {\bibfnamefont {P.~M.}\ \bibnamefont
  {Zaretzki}},\ }\bibfield  {title} {\bibinfo {title} {Simultaneous
  approximation and interpolation with norm preservation},\ }\href
  {https://doi.org/https://doi.org/10.1016/0021-9045(71)90039-6} {\bibfield
  {journal} {\bibinfo  {journal} {J. Approx. Theory}\ }\textbf {\bibinfo
  {volume} {4}},\ \bibinfo {pages} {54} (\bibinfo {year} {1971})}\BibitemShut
  {NoStop}%
\bibitem [{\citenamefont {Deutsch}\ and\ \citenamefont
  {Morris}(1969)}]{deutsch-sain}%
  \BibitemOpen
  \bibfield  {author} {\bibinfo {author} {\bibfnamefont {F.}~\bibnamefont
  {Deutsch}}\ and\ \bibinfo {author} {\bibfnamefont {P.~D.}\ \bibnamefont
  {Morris}},\ }\bibfield  {title} {\bibinfo {title} {On simultaneous
  approximation and interpolation which preserves the norm},\ }\href
  {https://doi.org/https://doi.org/10.1016/0021-9045(69)90004-5} {\bibfield
  {journal} {\bibinfo  {journal} {J. Approx. Theory}\ }\textbf {\bibinfo
  {volume} {2}},\ \bibinfo {pages} {355} (\bibinfo {year} {1969})}\BibitemShut
  {NoStop}%
\bibitem [{\citenamefont {Yamabe}(1950)}]{yamabe}%
  \BibitemOpen
  \bibfield  {author} {\bibinfo {author} {\bibfnamefont {H.}~\bibnamefont
  {Yamabe}},\ }\bibfield  {title} {\bibinfo {title} {On an extension of the
  helly's theorem},\ }\href@noop {} {\bibfield  {journal} {\bibinfo  {journal}
  {Osaka J. Math}\ }\textbf {\bibinfo {volume} {2}},\ \bibinfo {pages} {15}
  (\bibinfo {year} {1950})}\BibitemShut {NoStop}%
\bibitem [{\citenamefont {Beatson}(1977)}]{beatson}%
  \BibitemOpen
  \bibfield  {author} {\bibinfo {author} {\bibfnamefont {R.~K.}\ \bibnamefont
  {Beatson}},\ }\emph {\bibinfo {title} {Degree of Approximation Theorems for
  Approximation with Side Conditions}},\ \href@noop {} {Ph.D. thesis},\
  \bibinfo  {school} {University of Canterbury} (\bibinfo {year}
  {1977})\BibitemShut {NoStop}%
\bibitem [{\citenamefont {Remez}(1934)}]{remez}%
  \BibitemOpen
  \bibfield  {author} {\bibinfo {author} {\bibfnamefont {E.}~\bibnamefont
  {Remez}},\ }\bibfield  {title} {\bibinfo {title} {Sur le calcul effectif des
  polynomes dapproximation de tchebichef},\ }\href@noop {} {\bibfield
  {journal} {\bibinfo  {journal} {C. R. Acad. Sci. Paris}\ }\textbf {\bibinfo
  {volume} {199}},\ \bibinfo {pages} {337} (\bibinfo {year}
  {1934})}\BibitemShut {NoStop}%
\bibitem [{\citenamefont {Grenez}(1983)}]{grenez}%
  \BibitemOpen
  \bibfield  {author} {\bibinfo {author} {\bibfnamefont {F.}~\bibnamefont
  {Grenez}},\ }\bibfield  {title} {\bibinfo {title} {Design of linear or
  minimum-phase fir filters by constrained chebyshev approximation},\ }\href
  {https://doi.org/https://doi.org/10.1016/0165-1684(83)90091-9} {\bibfield
  {journal} {\bibinfo  {journal} {Signal Process.}\ }\textbf {\bibinfo {volume}
  {5}},\ \bibinfo {pages} {325} (\bibinfo {year} {1983})}\BibitemShut {NoStop}%
\bibitem [{\citenamefont {{Shor}}(2004)}]{shor_04}%
  \BibitemOpen
  \bibfield  {author} {\bibinfo {author} {\bibfnamefont {P.~W.}\ \bibnamefont
  {{Shor}}},\ }\bibfield  {title} {\bibinfo {title} {The adaptive classical
  capacity of a quantum channel, or information capacities of three symmetric
  pure states in three dimensions},\ }\href
  {https://doi.org/10.1147/rd.481.0115} {\bibfield  {journal} {\bibinfo
  {journal} {IBM Journal of Research and Development}\ }\textbf {\bibinfo
  {volume} {48}},\ \bibinfo {pages} {115} (\bibinfo {year} {2004})}\BibitemShut
  {NoStop}%
\bibitem [{\citenamefont {Peres}\ and\ \citenamefont
  {Wootters}(1991)}]{peres_wootters}%
  \BibitemOpen
  \bibfield  {author} {\bibinfo {author} {\bibfnamefont {A.}~\bibnamefont
  {Peres}}\ and\ \bibinfo {author} {\bibfnamefont {W.~K.}\ \bibnamefont
  {Wootters}},\ }\bibfield  {title} {\bibinfo {title} {Optimal detection of
  quantum information},\ }\href {https://doi.org/10.1103/PhysRevLett.66.1119}
  {\bibfield  {journal} {\bibinfo  {journal} {Phys. Rev. Lett.}\ }\textbf
  {\bibinfo {volume} {66}},\ \bibinfo {pages} {1119} (\bibinfo {year}
  {1991})}\BibitemShut {NoStop}%
\bibitem [{\citenamefont {Parvathalu}\ and\ \citenamefont
  {Johnson}(2017)}]{PJ_17}%
  \BibitemOpen
  \bibfield  {author} {\bibinfo {author} {\bibfnamefont {B.}~\bibnamefont
  {Parvathalu}}\ and\ \bibinfo {author} {\bibfnamefont {P.~S.}\ \bibnamefont
  {Johnson}},\ }\bibfield  {title} {\bibinfo {title} {Construction of
  mercedes–benz frame in $\mathbb{R}^n$},\ }\href
  {https://doi.org/10.1007/s40819-017-0367-8} {\bibfield  {journal} {\bibinfo
  {journal} {Int. J. Appl. Comput. Math}\ }\textbf {\bibinfo {volume} {3}},\
  \bibinfo {pages} {511} (\bibinfo {year} {2017})}\BibitemShut {NoStop}%
\bibitem [{\citenamefont {Mohammad-Abadi}\ and\ \citenamefont
  {Najafi}(2012)}]{MN_12}%
  \BibitemOpen
  \bibfield  {author} {\bibinfo {author} {\bibfnamefont {S.~A.}\ \bibnamefont
  {Mohammad-Abadi}}\ and\ \bibinfo {author} {\bibfnamefont {M.}~\bibnamefont
  {Najafi}},\ }\bibfield  {title} {\bibinfo {title} {Type of equiangular tight
  frames with $n + 1$ vectors in $\mathbb{R}^n$},\ }\href@noop {} {\bibfield
  {journal} {\bibinfo  {journal} {Int. J. Appl. Math. Res.}\ }\textbf {\bibinfo
  {volume} {1}},\ \bibinfo {pages} {391} (\bibinfo {year} {2012})}\BibitemShut
  {NoStop}%
\bibitem [{\citenamefont {Kitaev}\ \emph {et~al.}(2002)\citenamefont {Kitaev},
  \citenamefont {Shen},\ and\ \citenamefont {Vyalyi}}]{kitaev}%
  \BibitemOpen
  \bibfield  {author} {\bibinfo {author} {\bibfnamefont {A.~Y.}\ \bibnamefont
  {Kitaev}}, \bibinfo {author} {\bibfnamefont {A.~H.}\ \bibnamefont {Shen}},\
  and\ \bibinfo {author} {\bibfnamefont {M.~N.}\ \bibnamefont {Vyalyi}},\
  }\href@noop {} {\emph {\bibinfo {title} {Classical and Quantum
  Computation}}},\ \bibinfo {series} {Graduate Studies in Mathematics},
  Vol.~\bibinfo {volume} {47}\ (\bibinfo  {publisher} {American Mathematical
  Society},\ \bibinfo {year} {2002})\BibitemShut {NoStop}%
\bibitem [{\citenamefont {Nielsen}\ and\ \citenamefont
  {Chuang}(2011)}]{nielsen_chuang}%
  \BibitemOpen
  \bibfield  {author} {\bibinfo {author} {\bibfnamefont {M.~A.}\ \bibnamefont
  {Nielsen}}\ and\ \bibinfo {author} {\bibfnamefont {I.~L.}\ \bibnamefont
  {Chuang}},\ }\href@noop {} {\emph {\bibinfo {title} {Quantum Computation and
  Quantum Information: 10th Anniversary Edition}}},\ \bibinfo {edition} {10th}\
  ed.\ (\bibinfo  {publisher} {Cambridge University Press},\ \bibinfo {address}
  {USA},\ \bibinfo {year} {2011})\BibitemShut {NoStop}%
\bibitem [{\citenamefont {Childs}\ \emph {et~al.}(1999)\citenamefont {Childs},
  \citenamefont {Preskill},\ and\ \citenamefont {Renes}}]{childs}%
  \BibitemOpen
  \bibfield  {author} {\bibinfo {author} {\bibfnamefont {A.}~\bibnamefont
  {Childs}}, \bibinfo {author} {\bibfnamefont {J.}~\bibnamefont {Preskill}},\
  and\ \bibinfo {author} {\bibfnamefont {J.}~\bibnamefont {Renes}},\ }\bibfield
   {title} {\bibinfo {title} {Quantum information and precision measurement}}
  (\bibinfo {year} {1999}),\ \bibinfo {note} {arXiv Preprint.
  \url{https://arxiv.org/abs/quant-ph/9904021v2}}\BibitemShut {NoStop}%
\end{thebibliography}%

\end{document}